\documentclass[11pt]{article}
\newcommand{\remove}[1]{}
\usepackage{algorithm}
\usepackage{graphicx}
\usepackage{theorem}
\usepackage{epsf}
\usepackage{citesort}
\usepackage{ifthen}
\newboolean{marginnotes}
\setboolean{marginnotes}{false}
\usepackage{float}
\usepackage{amsmath}
\usepackage{amsfonts}
\usepackage{amsmath, amssymb}
\usepackage{epsfig}
\usepackage{subfigure}
\usepackage{graphicx}
\usepackage{epstopdf}
\newcounter{notecounter}

\ifthenelse{\boolean{marginnotes}}

{\newcommand{\note}[1]{\stepcounter{notecounter}%
\ifmmode{^{\alph{notecounter}}}\else{$^{\alph{notecounter}}$}\fi%
\marginpar{\footnotesize $^{\alph{notecounter}}$#1}} }

\def \ca{{\cal A}}

\newcommand{\bqa}{\begin{eqnarray}}
\newcommand{\eqa}{\end{eqnarray}}
\newcommand{\beq}{\begin{equation}}
\newcommand{\eeq}{\end{equation}}
\newcommand{\bqaa}{\begin{eqnarray*}}
\newcommand{\eqaa}{\end{eqnarray*}}

\newtheorem{theorem}{Theorem}[section]

\newtheorem{lemma}{Lemma}[section]

\newtheorem{definition}{Definition}[section]

\newenvironment{proof}{\indent {\bf Proof.}}{\hfill$\bf \Box$\bigskip}
\title{A Token Based Approach to Distributed Computation in Sensor Networks \thanks{This research
was supported by NSF Grant 0932114 and NSF CAREER awards ECS-0449194 and CNS-0238397.}}

\author{Venkatesh Saligrama, Murat Alanyali\\
Department of Electrical and Computer Engineering\\
Boston University, Boston, MA 02215, USA\\
Email: \{srv,alanyali\}@bu.edu}
\date{}
\begin{document}
\maketitle
\begin{abstract}
We consider distributed algorithms for data aggregation and function
computation in sensor networks. The algorithms perform pairwise computations along edges of an
underlying communication graph. A token is associated with each sensor node, which acts as a transmission permit. Nodes with active tokens have transmission permits; they generate messages at a constant rate and send each message to a randomly selected neighbor. By using different strategies to control the transmission permits we can obtain tradeoffs between message and time complexity. Gossip corresponds to the case when all nodes have permits all the time. We study algorithms where permits are revoked after transmission and restored upon reception. Examples of such algorithms include Simple-Random Walk(SRW), Coalescent-Random-Walk(CRW) and Controlled Flooding(CFLD) and their hybrid variants. SRW has a single node permit, which is passed on in the network. CRW, initially initially has a permit for each node but these permits are revoked gradually. The final result for SRW and CRW resides at a single(or few) random node(s) making a direct comparison with GOSSIP difficult. A hybrid two-phase algorithm switching from CRW to CFLD at a suitable pre-determined time can be employed to achieve consensus.  We show that such hybrid variants achieve significant gains in both message and time complexity. The per-node message complexity for n-node graphs, such as 2D mesh, torii, and Random geometric graphs, scales as $O(polylog(n))$ and the corresponding time complexity scales as $O(n)$. The reduced per-node message complexity leads to reduced energy utilization in sensor networks.
\end{abstract}

\section{Introduction}

Large sensor systems have remarkable potential in a wide range of applications from environmental monitoring to intrusion detection. Such systems are now viable thanks to recent progress in integration and communication technologies, yet their size precludes classical telemetry to collect sensory data and poses algorithmic challenges in handling the high data volume. The principle of gossip offers an appealing and scalable solution approach to this issue: In broad terms, gossip algorithms are decentralized methods to compute statistics of system-wide data based on asynchronous message passing between sensors that are within immediate communication range.  The purpose of this paper is to introduce and analyze token-based gossip algorithms, and put them in perspective with previously studied gossip algorithms with respect to computation time, energy consumption, accuracy, and robustness.

The generic gossip algorithm has been well-studied in the context of computing averages~\cite{kempe:2003,Boyd:2005,Scherber:2005}, and has roots in load balancing~(see, for example,~\cite[Section7.4]{Tsitsiklis89}).  Here it is assumed that each sensor has a local scalar value and it is of interest to compute the average of these values.  Gossip algorithms accomplish this task by randomly choosing two neighboring sensor nodes at each time and replacing their current values by their average. It turns out that under mild conditions this process over time converges to the average of all sensor values at all the sensors, i.e., the sensors asymptotically achieve a consensus. The algorithm executes autonomously at each sensor; it is robust to communication errors and sensor failures; and a final state of consensus provides robustness and convenience in reading the average from the system. Such consensus algorithms have been recently explored in other contexts such as detection~\cite{acc2004,Saligrama:2005}. 

Nevertheless, these algorithms have fundamental disadvantages from an energy efficiency perspective. For example it is well known that in grids and in tori with $n$ nodes, the number of message transmissions {\em per node} to complete the computational task scales as $\Omega(n)$~\cite{Boyd:2005,SavasCISS}, which can be significant for a large sensor network. The fundamental reason is that energy
efficiency resulting from in-network processing is offset by ad-hoc
message passing that results in redundant computations, i.e., the
same set (or largely similar set) of nodes repeatedly fuse their
information at different points in time. In a related problem
involving distributed detection, the significant energy scaling can
be attributed to the loopy nature of the network where messages sent
from one node repeatedly arrive at the node at different points in
time. In order to ensure that no information from any node is
forgotten, each node must re-inject its value into the network
to reinforce its information~\cite{Saligrama:2005}. At a
fundamental level the significant scaling of energy arises due to
the slow mixing rate of large networks, which can be attributed to
rather large second eigenvalues of certain connectivity matrices
associated with the underlying communication graph~\cite{Boyd:2005}[Theorem~3].

Another important disadvantage of generic gossip algorithms concerns accuracy. Stopping criterion of gossip is based on tail probabilities of a normalized distance between current system state and state of consensus~\cite{Boyd:2005}. In turn, these algorithms provide only probabilistic guarantees on final consensus and therefore they should be considered as approximation algorithms. It should also be noted that even when such a guarantee holds and the final normalized error happens to be small, the actual error may be substantial if the normalizing constant is large.  Such situations arise in large networks in which a  small set of sensor values significantly influence the sought value.

To put the present paper in perspective with gossip algorithms it helps to consider signal processing and communications {\em separately}. Here we adopt  the objective of {\em exact} computation of a function of distributed data, and consider performance of a novel class of communication algorithms towards that end. In this view conventional gossip may be considered as a communication algorithm to {\em approximate} the same function. This approximation is clearly one specific instantiation; and there are other wide range of approximation criteria that may be useful to consider from a signal processing perspective. From a signal processing point of view it also makes sense to incorporate prior signal information, such as boundedness, distribution of values etc. While these issues are important our focus here is primarily on the communication aspects for exact distributed computation.

\subsection{Token-Based Algorithms}
We present a novel concept of token-based gossip for distributed computing. This concept preserves the ad-hoc network operation but entails perfect accuracy of computation and exponential savings in energy-consumption over the existing local message passing algorithms. Under the algorithms studied here, a transmitting node becomes inactive and does not transmit further messages until it is reactivated by a message reception from another node. An active node generates messages at constant rate and sends each message to a randomly selected neighbor. Hence network nodes implement interactive sleep-wake schedules. Active nodes are interpreted to hold imaginary tokens that act as transmission permits. The total energy consumption is controlled by managing the number of tokens in the network.

\begin{figure}[htbp]
\centerline{\includegraphics[height=3in]{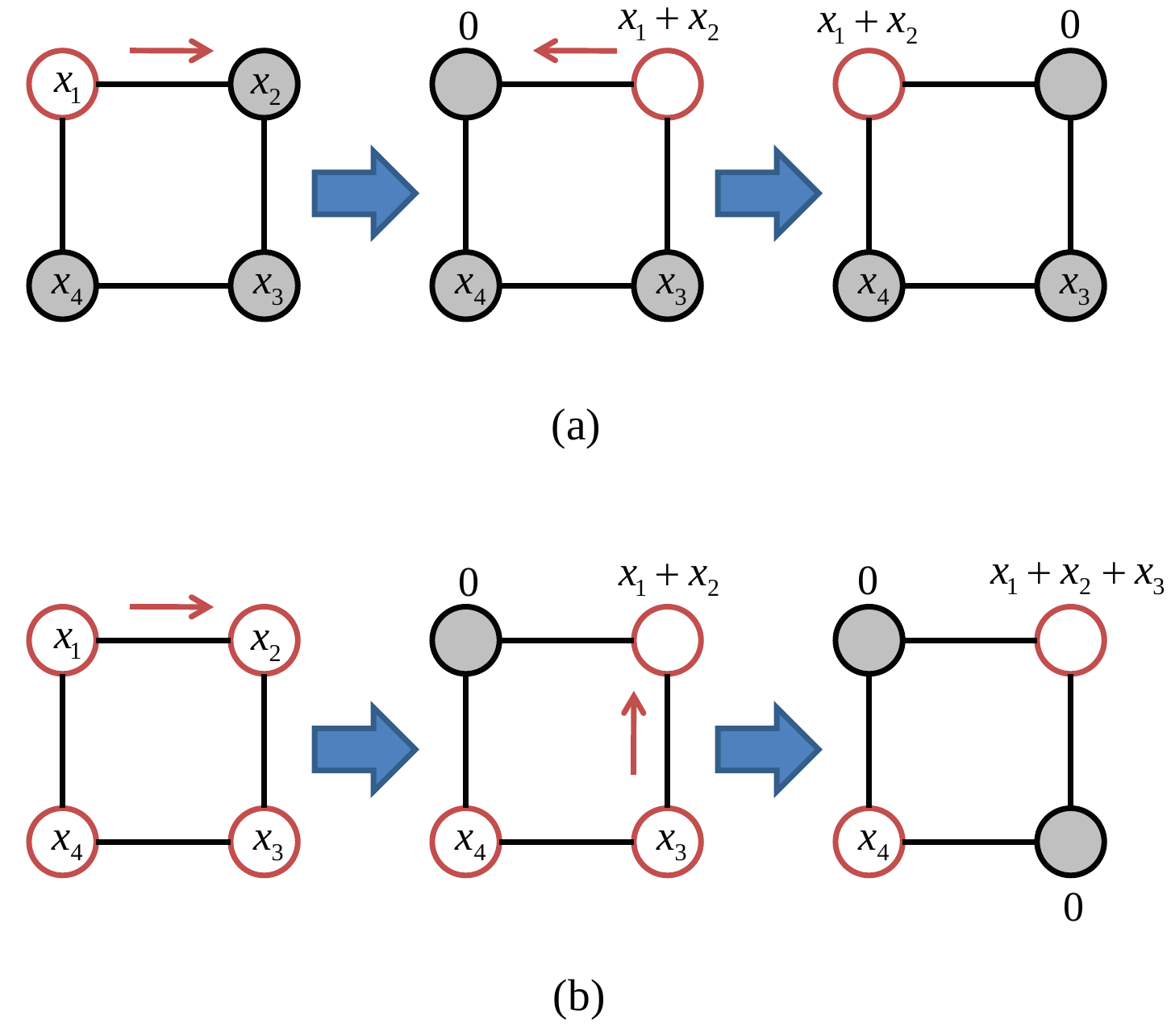}}
\caption{\small Illustration of token-based algorithms: (a) SRW and (b) CRW. Each node is either active or inactive. Active nodes are indicated with light color. Nodes can transmit information in the active state but not in the inactive state. A node can transition from active to inactive or vice versa based on pre-defined protocol. In SRW there is always one active node. In CRW every node is active initially but the number of active nodes decrease in time, towards the final value one.}
\label{fig.algorithms}
\end{figure}

We describe different instances of token based algorithms: (i)  Algorithm {SRW} maintains a single active node (i.e.~a single token), whose trajectory is a random walk on the communication graph. Local processing at each active node exploits a decomposability property  of the considered function to guarantee that the function is computed when each node becomes active at least once. In turn performance of this algorithm is closely related to the cover time~\cite{Zuckerman:1992} of random walks.  (ii) Under algorithm {CRW} all nodes are initially active but when two active nodes communicate with each other their tokens coalesce and therefore the number of tokens in the system reduces by one.  The computation is completed when a single token remains in the system. This latter algorithm is closely related with coalescing random walks~\cite{Cox:1989} that have been studied as duals of a class of interacting particle systems coined as voter models. The two algorithms are illustrated in Figures~\ref{fig.algorithms}(a)-(b).
\begin{figure}[htbp]
\centerline{\includegraphics[height=3in]{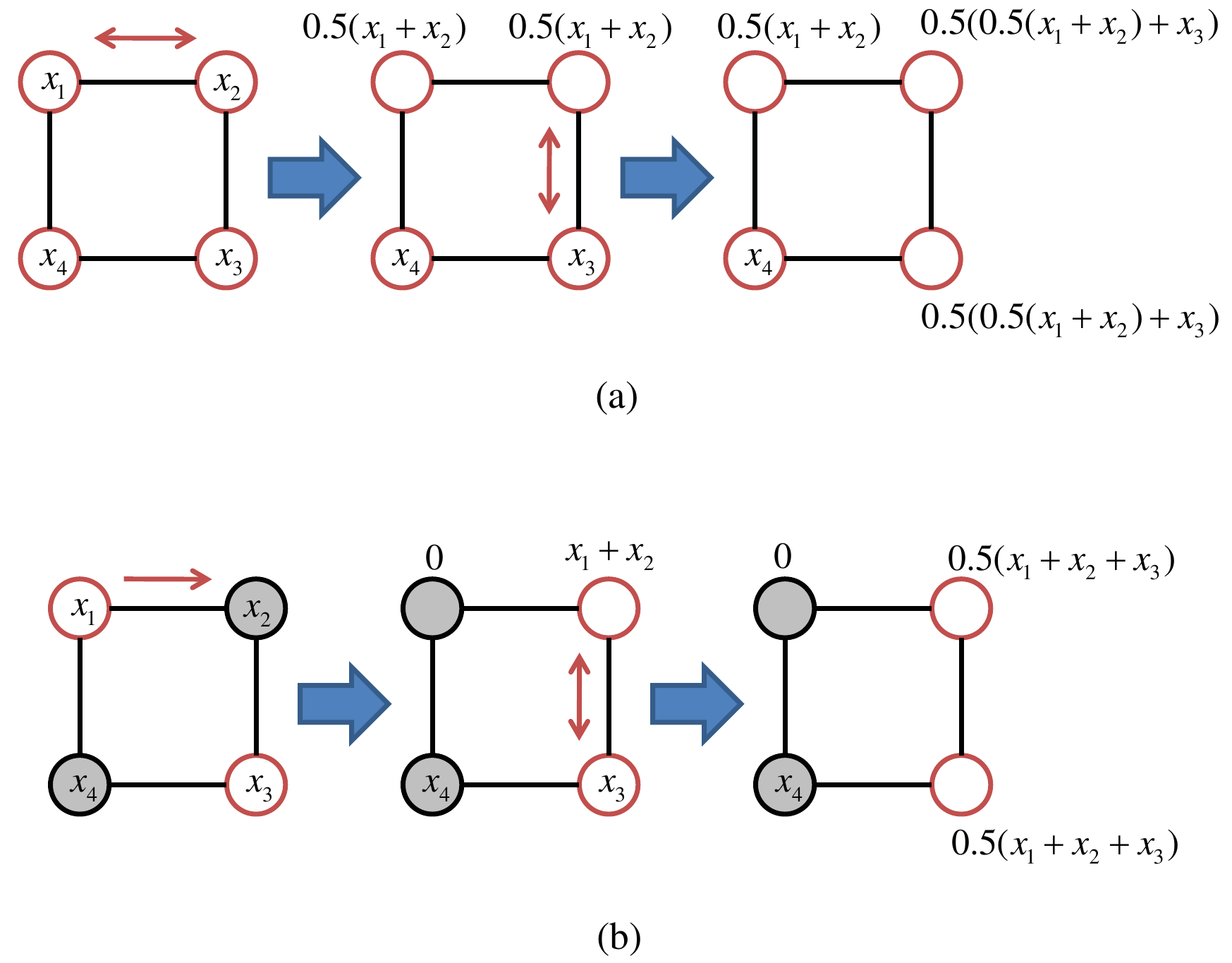}}
\caption{\small Illustration of two algorithms in the framework of the paper. (a) In conventional gossip all nodes are active all the time. Transmission is two-way and two nodes that share information replace their prior values with the fused values. (b) Hybrid token algorithm that maintains a constant (in this case 2) number of active nodes.  Transactions between an active and an inactive node are governed by the rules of SRW, while transactions between two active nodes are governed by rules of conventional gossip.}
\label{fig.variations}
\end{figure}
A range of distributed algorithms can be obtained by variations of the token-based communication concept. For example, conventional gossip can be considered as a token-based algorithm where each node maintains a token at all times. The local processing upon each message exchange in this case is illustrated in Figure~\ref{fig.variations}(a). Alternatively, one may consider hybrid schemes to improve message and time complexity. One hybrid scheme involves with a fixed but arbitrary number of tokens. Tokens progress from an active node to an inactive node while updating local values exactly as in SRW. If two active nodes interact then they both relax their values as in conventional gossip, and remain active.  An illustration of such a scheme is given in Figure~\ref{fig.variations}(b). Another hybrid scheme involves switching at some time $t$, from CRW to Controlled Flooding (CFLD)~\cite{bertsekas}, which is a token based algorithm where tokens multiply at each local broadcast. 
In contrast to flooding, CFLD follows additional rules to control the number of transmissions(see Section~\ref{sec.twophase}).

\subsection{Time and Message Complexities}
In this section we describe tradeoffs between time and message complexities for regular graphs to illustrate some of the benefits of SRW and CRW. However, a fundamental difference between Gossip algorithm and SRW/CRW makes this comparison potentially difficult. Note that in the standard Gossip algorithm, the fused value is a consensus estimate at all the nodes. In contrast SRW/CRW realize the fused value at a random node in the network. \emph{In practice it is desirable to have access to the fused value at a designated node(s).} Consequently, to meet this requirement, SRW/CRW would have to transmit the fused estimate to the designated fusion center. This raises two fundamental issues: \\
{\bf (A)} How can a node recognize that it has the fused estimate? \\
{\bf (B)} How to efficiently transmit the fused information to the designated fusion center? \\
As it turns out, both of these questions can be addressed satisfactorily. To address the first issue we augment the distributed computation problem with a secondary distributed computation procedure. This secondary computation determines when fusion has been realized. To transmit this information to the designated node(s) we \emph{flood} the entire network through CFLD. The overall message complexity for CFLD for a single message scales as the number of communication links in the network. The time complexity scales as the diameter of the network~\cite{heinz}. Furthermore, other choices can result in superior performance. For instance, nodes follow the CRW protocol until a predesignated time $t$ and switch to CFLD after time $t$.

Table~\ref{table.complexity} (see Section~\ref{sec.toruscomp} and \ref{sec.arbitrary}) illustrates completion times and per-node transmission counts for regular topologies.  For the $d$-dimensional lattice torus with $n$ sensors, we the completion time of CRW is
$\Theta(n(\log n)^\alpha)$ and energy requirement per sensor node is $O((\log n)^{\alpha+1})$ where $\alpha=1$ for $d=2$ and $\alpha=0$
for $d\ge 3$. The algorithm thus has a favorable energy scaling, furthermore its performance is almost insensitive to changes in the
network connectivity represented by different values of $d\ge 2$, hinting at the possibility of predictable performance over mesh
topologies. Both the time and the per-node energy requirement of the generic gossip algorithm of~\cite{Boyd:2005} scale as
$\Omega(n)$ on the 2-dimensional torus with $n$ nodes. We also depict results for the two phase CRW + CFLD algorithm. For the 2D torus the time complexity is similar to GOSSIP and the message complexity is similar to CRW. Therefore, this two-phase scheme is an improvement over both GOSSIP and CRW. Recent results indicate that the energy scaling can be improved significantly via variations of gossip that require location awareness capability for each sensor~\cite{Dimakis:2006,Li:2007}. Similar variations of token-based gossip may also yield reduction in energy complexity, though that direction is not pursued in this paper.

\begin{table}\centering
\begin{tabular}{c|c|c|c|c}
& CRW & SRW & GOSSIP & CRW + CFLD\\ \hline
Clique&$\Theta(\log n)$&$\Theta(\log n)$& $\Theta(n)$& $\Theta(\log n)$\\
Ring&$\Theta(n)$&$\Theta(n)$& $\Theta(n^2)$& $\Theta(n)$\\
Torus~($d=2$)&$O(\log^2 n)$&$\Theta(\log^2 n)$&$\Theta(n)$ & $O(\log^2 n)$\\
Torus~($d \geq 3$)&$O(\log n)$&$\Theta(\log n)$&$\Theta(n^{2/d})$ & $O(d+\log n)$\\
\end{tabular}\vspace*{0.4cm}

(a)\vspace*{0.6cm}

\begin{tabular}{c|c|c|c|c} &CRW&SRW&GOSSIP&CRW + CFLD\\ \hline
Clique&$\Theta(n)$&$\Theta(n\log n)$& $\Theta(n)$& $\Theta(n)$\\
Ring&$\Theta(n)$&$\Theta(n\log n)$&$\Theta(n^2)$& $\Theta(n)$\\
Torus~($d=2$)&$\Theta(n\log n)$&$\Theta(n\log^2 n)$&$\Theta(n)$& $\Theta(n)$ \\
Torus~($d \geq 3$)&$\Theta(n)$&$\Theta(n\log n)$&$\Theta(n^{2/d})$& $\Theta(n)$ \\
\end{tabular}\vspace*{0.4cm}

(b)

\caption{(a) Message and (b) time complexities
with $n$ nodes.} \label{table.complexity}
\end{table}


The conclusions of Table~\ref{table.complexity} (see Sections~\ref{sec.toruscomp} and \ref{sec.arbitrary}) for regular topologies is presented in Section~\ref{sec.torus} and is largely based on existing results in applied probability literature. Preliminary work along these lines has also been described by the authors~\cite{SavasCISS}. This paper develops results for general topologies based on a deeper analysis of hitting-time computations on the communication graph. While our approach can lead to conservative estimates for general graphs, it turns out that for graphs with local neighborhood structure, such as grids and random geometric graphs(RGG), these estimates are relatively tighter. Our message complexity for RGG scales as $O(\log^2(n))$ and can be combined with CFLD to realize consensus with significant improvement in message and time complexity over GOSSIP.

\noindent
{\bf Robustness:} Similar to other token-based communication algorithms such as token rings,  robustness of the introduced algorithms suffers from the potential of losing tokens.  Packet losses do not contribute to this potential if reliable link protocols are adopted.  However permanent node failures may have an impact on system performance if a node fails while holding a token. The issue may be mitigated by running multiple independent instances of a token-based algorithm simultaneously, in order to reduce the likelihood of losing all tokens simultaneously at a failing node, without altering the scaling of time and message complexities. As will be clear in the sequel, a token-bearing node knows explicitly the number of sensor values fused in its current value; and that value may be useful partial information in case of node failures.  Alternatively a hybrid scheme with multiple tokens may be invoked, thereby providing a robustness akin to conventional gossip.

\noindent
{\bf Paper Organization:} The paper is organized as follows. In Section~\ref{sec.problem} we formalize a general distributed computation problem that includes computation of statistics of spatially dispersed sensory data. The two token-based gossip algorithms SRW and CRW are formally specified in Section~\ref{sec.algo} and their correctness is established. Section~\ref{sec.torus} illustrates time and message complexities for regular topologies as summarized by Table~\ref{table.complexity}. Section~\ref{sec.twophase} describes two-phase algorithms that combine CRW with CFLD to realize a consensus fused estimate. Section~\ref{sec.arbitrary} gives a novel analysis of coalescing random walks on arbitrary graphs and thereby determines the complexity of CRW in the general setting.

\section{Problem formulation}\label{sec.problem}

We consider a collection of $n$ networked sensors. The communication graph of this collection is an undirected graph $G=(V,E)$ in which each sensor is uniquely represented by a node in~$V$. An edge in $E$  indicates that the two sensors corresponding to its incident nodes have a bidirectional communication link.  In order to avoid trivialities $G$ is assumed  to be connected; otherwise it is arbitrary.

Each sensor $i$ has a value $x_i$ and we are interested in computing a function $F_n(x_1,x_2,\cdots,x_n)$ of the $n$ sensor values. The function $F_n(\cdot)$ is assumed to be symmetric so that for any permutation $\pi$ of $\{1,2,\cdots,n\}$
\[F_n(x_1,x_2,\cdots,x_n)=F_{n}(x_{\pi_1},x_{\pi_2},\cdots,x_{\pi_n}).\] Furthermore we assume existence of an atomic function $f(\cdot)$ such that for each $1\le k\le n$
\begin{equation} F_n(x_1,x_2,\cdots,x_n)= f(F_k(x_{\pi_1},\cdots,x_{\pi_k}),
F_{n-k}(x_{\pi_{k+1}},x_{\pi_{k+2}},\cdots,x_{\pi_n})).\label{eq.atom1} \end{equation}
In particular \begin{equation} f(x_i,x_j)=F_2(x_i,x_j).\label{eq.atom2}
\end{equation}

For example if $f(x_i,x_j)=\max(x_i,x_j)$ or if  $f(x_i,x_j)=x_i+x_j$ then $F_n(\cdot)$ is  respectively the maximum or the sum of $x_1,x_2,\cdots,x_n$.  If each $x_i=(y_i,w_i)$ is a vector and $f(\cdot)$ is the vector-valued function \[f(x_i,x_j)=((w_i+w_j)^{-1}(w_iy_i+w_jy_j)\;,\;w_i+w_j),\] then $F_n(\cdot)$ is the tuple \[F_n(x_1,x_2,\cdots,x_n)=\left(\sum_{i=1}^n\frac{w_i}{\sum_{j=1}^nw_j}y_i~,~\sum_{i=1}^nw_i\right).\]
Weighted average computations of this sort are particularly relevant to applications of Kalman filtering in distributed tracking~\cite{Rahman:2007}.

We finally assume existence of a special value $e$ that acts as an identity element for $f(\cdot)$ so that for any value $x$  \[f(x,e)=f(e,x)=f(e,e)=e.\] This element is not a fundamental requirement, and in fact it makes the upcoming algorithm specifications look somewhat mysterious, but it is useful in giving a concise reasoning about correctness of the algorithms introduced next.

\section{Token-based Gossip Algorithms}\label{sec.algo}

We specify two algorithms, namely SRW and CRW. A pseudo-code for these
algorithms is given in Figure~\ref{fig.PCode}. Under each algorithm, each node maintains three variables \texttt{value},
\texttt{status}, and {\tt count}. Content of \texttt{value} is an estimate of $F_n(x_1,x_2,\cdots.x_n)$. Initially {\tt value}$=x_i$ and {\tt count}$=1$ at each node $i$. The variable \texttt{status} is either \emph{`active'} or \emph{`inactive'} and it indicates whether the node is holding a token or not. Let \begin{eqnarray*}
v_i(t)&=&\mbox{ content of {\tt value} at node $i$ at time $t$,}\\
\xi_i(t)&=&\left\{\begin{array}{ll}1&\mbox{ if {\tt status} of node
$i$ is `active' at time $t$},\\0&\mbox{ else.}
\end{array}\right. \\
c_i(t)&=&\mbox{ content of {\tt count} at node $i$ at time $t$}.
\end{eqnarray*}

Hence  $v_i(0)=x_i$ and $c_i(0)=1$ for each node $i$. The initial value
$\xi_i(0)$ (i.e.~of {\tt status}) depends on the particular
algorithm: Under {SRW}
$\xi_i(0)=1$ for exactly one node, say node~$i_o$, whereas under CRW
$\xi_i(0)=1$ for all nodes~$i$.

\begin{algorithm}[t]
\centerline{\begin{minipage}{2.5in}
Variables: {\tt status, value, count}.\\
{\bf Initialize}: {\tt value} $\leftarrow x_i$; {\tt count} $\leftarrow$ 1;
\[\begin{array}{ll} \mbox{ SRW}:&{\tt status} \leftarrow
\left\{\begin{array}{ll}\mbox{`active'}& \mbox{ if }
i=i_o;\\\mbox{`inactive'}&\mbox{ else}.\end{array}\right.\\
\mbox{ CRW}:&{\tt status} \leftarrow \mbox{ `active'}.
\end{array}\]
\begin{minipage}[t]{2.5in}\begin{tabbing}
P\=rocedure {\bf Send}()\\
\> if\= ( {\tt status} $==$ `active' ) $\{$\\
\> \>choose neighbor;\\
\> \>send(neighbor , {\tt value, count});\\
\> \>{\tt value} $\leftarrow e$;\\
\>\> {\tt count} $\leftarrow 0$;\\
\> \>{\tt status} $\leftarrow$ `inactive';\\
\>\>$\}$
\end{tabbing}\end{minipage}  \\ \\
\begin{minipage}[t]{2.5in}\begin{tabbing}
P\=rocedure {\bf Receive}( {\tt value\_in, count\_in}) \{\\
\> {\tt value} $\leftarrow f$({\tt value, value\_in});\\
\>{\tt count} $\leftarrow$ {\tt count} $+$ {\tt count\_in};\\
\> {\tt status} $\leftarrow$ `active';\\
\>\}
\end{tabbing}\end{minipage}
\end{minipage}
} \caption{\small Pseudo-code for algorithms { SRW} and {
CRW} at node $i$. Send() is activated by the local Poisson
clock at the node, and Receive() is activated by message reception
from some other node. The two algorithms differ in the initialization
of the variable {\tt status}. Each algorithm terminates when {\tt count} is equal to the number of nodes in system. }\label{fig.PCode}
\end{algorithm}

Variables {\tt value, status} evolve according to the same rules under both algorithms: Namely, each node has an independent Poisson clock that ticks at unit rate. When the local clock of a node ticks, the node does not take any action unless it is active at that time. If the node is active, then it chooses a neighbor at random, sends its current {\tt value} and {\tt count} to that neighbor, and becomes inactive. The Send() subroutine of the algorithm maintains {\tt value} and carries the identity $e$. The {\tt count} is 0 at each inactive node at all times. If the selected neighbor was inactive at the time of reception, then it simply adopts the variables of the sender and it activates itself. Otherwise, the node executes a step towards computation of $F_n$ and adds the received {\tt count} value to its own.

For each time $t$ define
$\xi(t)\triangleq(\xi_1(t),\xi_2(t),\cdots,\xi_n(t))$. The process $(\xi_1(t),\xi_2(t),\cdots,\xi_n(t))$ of activity indicators is Markovian due to the randomness of the choice of neighbor by active nodes: For SRW the unique 1 in $(\xi_1(t),\xi_2(t),\cdots,\xi_n(t))$ follows a simple random walk on the communication graph. For CRW $(\xi_1(t),\xi_2(t),\cdots,\xi_n(t))$ indicates sites occupied by $n$ simple random walks each of which evolves independently until it meets another and makes identically the same transitions with that walk afterwards.

\subsection{Algorithm Correctness}

We first establish correctness of the introduced algorithms by showing that each algorithm computes the quantity $F_n(x_1,x_2,\cdots,x_n)$ in finite time.

\begin{lemma}\label{lemma.invariant}
Under both { SRW} and { CRW}, and for all $t>0$,
\[F_n(v_1(t),v_2(t),\cdots,v_n(t))=F_n(x_1,x_2,\cdots,x_n).\]
\end{lemma}

\begin{proof}
We prove the lemma by induction. Let $t_0=0$ and let $t_k$ be the time of $k$th message passing in the system.  The claim is true at time $t_0$ due to the initialization of each algorithm. Since the system state remains constant in the interval $(t_k,t_{k+1})$, it is enough to show the claim holds at time $t_{k+1}$ provided that it holds at time $t_k$. To this end suppose the claim holds at time $t_k$. Without loss of generality, suppose that the $k+1$st message has sender~1 and receiver~2. Then $v_1(t_{k+1})=e,~v_2(t_{k+1})=f(v_1(t_k),v_2(t_k))$ and $v_l(t_{k+1})=v_l(t_k)$ for $3\le l\le n$. Therefore
\bqaa
F_n(v_1(t_{k+1}),v_2(t_{k+1}),\cdots,v_n(t_{k+1}))&=&F_n(e,f(v_1(t_k),v_2(t_k)),v_3(t_k),\cdots,v_n(t_k))\\
&=&f(~F_2(e,f(v_1(t_k),v_2(t_k)))~,~F_{n-2}(v_3(t_k),\cdots,v_n(t_k))~)\\
&=&f(~F_2(v_1(t_k),v_2(t_k))~,~F_{n-2}(v_3(t_k),\cdots,v_n(t_k))~)\\
&=&F_n(v_1(t_k),v_2(t_k),\cdots,v_n(t_k)),
\eqaa where the second and fourth equalities are due to (\ref{eq.atom1}) and the third equality is due to (\ref{eq.atom2}). This establishes the induction step and in turn the desired conclusion.
\end{proof}

We first consider SRW and define
\[\tau_S=\inf\{t:\mbox{ each node becomes active at least once by time $t$ }\}.\]
Let $a(t)$ denote the node that is active at time $t\ge \tau_S$. Since each node has been active at least once by time $t$, every node other than $a(t)$ should have turned inactive by sending the token to a neighbor. Therefore at each node $i\ne a(t)$ the value is $v_i(t)=e$. From Equations~(\ref{eq.atom1})--(\ref{eq.atom2}) we get \[F_n(v_1(t),\cdots,v_n(t))~=~F_2(v_{i(t)}(t),e)~=~v_{i(t)}(t).\] Therefore by Lemma~\ref{lemma.invariant} the value $v_{a(t)}(t)$ of node $a(t)$ is equal to $F_n(x_1,\cdots,x_n)$.

The above argument applies verbatim to CRW also, by taking $a(t)$ to be the unique active node at time $t\ge \tau_C$ where $\tau_C$ is defined as \[\tau_C=\inf\{t:\sum_i\xi_i(t)=1\}.\]

It should be clear that for SRW $\tau_S$ is the cover time of the communication graph $G$ and it is finite with probability~1. Under CRW,  $\sum_i\xi_i(t)$ is a non-increasing process with an absorption state at 1. In this case $\tau_C$ is the absorption time of this process and it is almost surely finite. Hence under each algorithm the desired quantity $F_n(x_1,\cdots,x_n)$ is available at some node within finite time.

It remains to establish how the termination times $\tau_S, \tau_C$ can be recognized. Towards this end note that the update mechanism for variable {\tt count} reflects a secondary distributed computation procedure with $f(c_i,c_j)=c_i+c_j$ and \[F_n(c_1(0),\cdots,c_n(0))~=~\sum_ic_i(0)~=~n.\] The general conclusions obtained above are valid in this special case, and in turn $c_{i(t)}(t)=n$ for $t\ge \tau_S$ under SRW and for $t\ge \tau_C$ under CRW. Therefore at such time instants node $i$ can identify itself as the unique bearer of the desired quantity by verifying the condition $c_i(t)=n$. In other words variable {\tt count} keeps an account of how many sensor values have been fused so far to form the content of variable {\tt value}; this serves as a pilot signal with a known terminal value that signals the end of each algorithm. We collect these observations in the following theorem:

\begin{theorem} \label{thm.correct}
Both SRW and CRW compute the exact value of $F_n(x_1,\cdots,x_n)$ in finite time. Each algorithm terminates with the correct value when the content of variable {\tt count} reaches $n$, the system size, at some node in the system.
\end{theorem}

\subsection{Time and Message Complexities}\label{sec.torus}

We present execution time and message complexity for SRW and CRW. We begin this section with the definitions of message and time complexities.

\begin{definition}
{\em Average time complexity} of {SRW}
(resp.~{CRW}) refers to $E[\tau_S]$ (resp.~$E[\tau_C]$).
\end{definition}

In adopting a measure of messaging complexity, let $\eta_{S}(t)$ and
$\eta_{C}(t)$ be the total number of transmitted messages in the
network by time $t$ under algorithms { SRW} and {
CRW} respectively:

\begin{definition}
{\em Average per-node message complexity} of  {
SRW} (resp.~{CRW}) refers to
$n^{-1}E[\eta_{S}(\tau_S)]$ (resp.~$n^{-1}E[\eta_{C}(\tau_C)]$).
\end{definition}

Note that
$(\xi(t):t\ge 0)$ is a Markov process under both
algorithms. More precisely, $(\xi(t):t\ge 0)$ is a random walk on
$G$ under {SRW}, and a coalescing random walk on $G$
under {CRW}. In particular the average time complexity of
{SRW} is the mean cover time of $G$. Average message
complexities of the algorithms are characterized by the following
lemma:

\begin{lemma}
\begin{eqnarray*}
E[\eta_{S}(\tau_S)]&=&E\left[\int_0^{\tau_S}\sum_i\xi_i(t)dt\right]~~=~~E[\tau_S],\\
E[\eta_{C}(\tau_C)]&=&E\left[\int_0^{\tau_C}\sum_i\xi_i(t)dt\right].
\end{eqnarray*} \label{lem:av_mes}
\end{lemma}
\begin{proof}
We provide a proof that applies verbatim to both algorithms.
Let $\eta(t)$ represent $\eta_S(t)$ for SRW and $\eta_C(t)$ for CRW.
Let $\mathcal{F}_t$ denote the sigma-field generated by
$(\xi(s):s\le t)$ and let $(\phi(t):t\ge 0)$ be a Poisson process
with unit rate. Note that $\eta(t)$ has the same distribution as
$\phi(\int_0^t\sum_i\xi_i(s)ds)$ since each carrier node transmits
messages at unit rate and inactive nodes do not engage in message
transmission, and thus $\sum_i\xi_i(t)$ is the instantaneous rate of
message generation in the network at time $t$. In particular the
process $(\mu(t):t\ge 0)$ with
\beq \mu(t)=\eta(t)-\int_0^t\sum_i\xi_i(s)ds \label{eq.lemmaaux1} \eeq is a
martingale adapted to $\{\mathcal{F}_t\}$. Both
$\tau_S$ and $\tau_C$ are $\{\mathcal{F}_t\}$-stopping times and
they are almost surely finite. In addition $\sup_{t\ge 0}\sum_i\xi_i(t)\le n$;
hence it follows by the optional sampling theorem~\cite[Theorem 2.2.13]{EthierKurtz} that
$E[\mu(\tau_S)]=E[\mu(\tau_C)]=0$. Using this observation in equality (\ref{eq.lemmaaux1}) establishes the lemma.
\end{proof}
\section{Regular Topologies} \label{sec.toruscomp}
Since the two algorithms are closely related to random walks, their time and message complexities for special topologies of the communication graph $G$ can be deduced by referring to related work in applied probability. Before giving an analysis for general topologies in the next section, we consider here the cases when $G$ is completely connected (i.e.~clique) and $d$-dimensional torus for $d\ge 1$. Recall that average time complexity of {SRW} is the mean cover time of $G$ and by the above lemma this is proportionally related to the mean message complexity. We refer the reader to~\cite[Chapter 5]{aldous:book} for the cover time results, which are summarized in Table~\ref{table.complexity} in the column SRW. We articulate on the time/message complexity of CRW in more detail, as that entails consideration of coalescing random walks, which are relatively obscure in engineering applications.

\paragraph{Completely connected graph:} A completely connected
graph is a graph where each vertex has an edge with every other
vertex.  In such graphs the process $(\sum_i\xi_i(t) \mid t>0)$ is also Markovian. This follows from the fact that the transition matrix at any time instant is invariant to permutation. For CRW this process has initial state $\sum_i\xi_i(0)=n$ and
at time $t<\tau_C$ it decreases by one at instantaneous rate
\[\left(\begin{array}{c}\sum_i\xi_i(t)\\2\end{array}\right),\]
that  is, the number of edges that are incident on two actives nodes at time $t$.
This process has been studied in detail in~\cite{Tavare:1984} and the
results therein are summarized in Table~\ref{table.complexity}.
While a completely connected graph reflects a limited set of cases
of practical importance, its analysis interestingly sheds considerable
light on mesh-type topologies that are considered next.

\paragraph{Ring and $d$-dimensional torus:} A $d$-dimensional torus
is a graph where all the vertices have exactly $2d$ neighbors and it
can be formed by joining the facing boundaries of a grid hence
yielding a completely symmetric structure. A ring is simply a 1-dimensional
torus. Consider $n = N^d$ for a $d$-dimensional torus. Let
\begin{eqnarray*}
s_N&=&\left\{\begin{array}{ll}
N^2 & \mbox{ if } d = 1 \\
N^2\log N & \mbox{ if }d = 2 \\
N^d & \mbox{ if }d \geq 3.
\end{array} \right.
\end{eqnarray*}

An asymptotic analysis of coalescing random walks on $d$-dimensional torus for large $N$ is given by Cox~\cite{Cox:1989}. It is established that the time-scaled the process $\sum_i\xi_i(s_Nt)$ on a $d$-dimensional torus converges in distribution to $\sum_i\xi_i(t)$ on a completely connected graph as $N\rightarrow\infty$.  (Note that $\sum_i\xi_i(t)$ is not even Markovian unless $G$ is completely connected.)  This result in turn leads to the following theorem on the time complexity for {CRW}:

\begin{theorem}\label{thm.timeComplexityC} \cite[Theorem 6]{Cox:1989}
\[0~<~\liminf_{N\rightarrow\infty}E[\tau_C]/s_N
~=~\limsup_{N\rightarrow\infty}E[\tau_C]/s_N~<~\infty.\]
\end{theorem}

\begin{figure}[t]\vspace*{-0.5cm}
\begin{center}\hspace*{-0.5cm}\epsfig{file=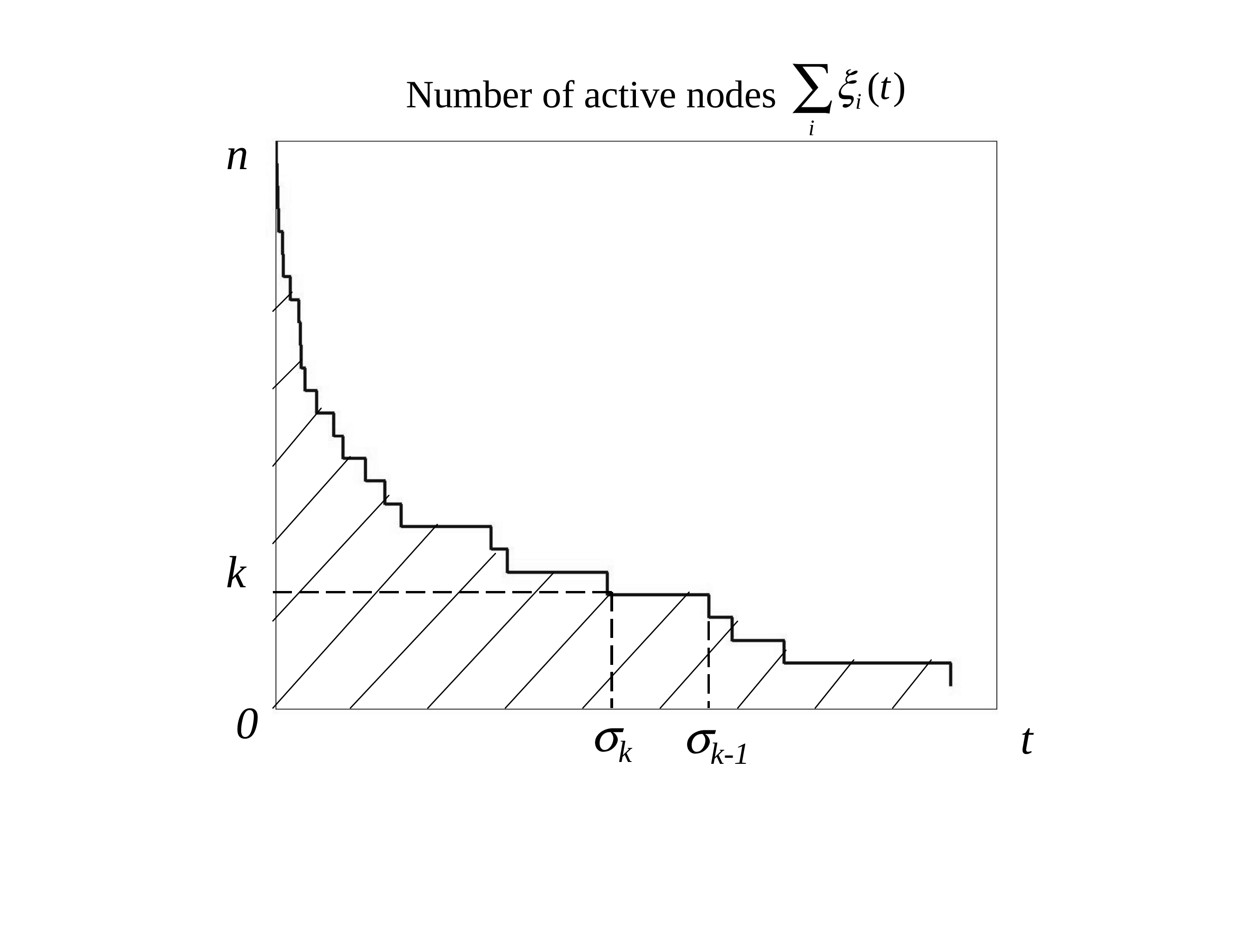, height=8cm}\end{center}
\vspace*{-1.5cm}
\caption{\small A sample path of the number of active nodes
under algorithm {CRW}. $\sigma_k$ denotes the first time
that $k$ active nodes remain in the network so that $\tau_C=\sigma_1$.
The mean of the shaded area is the mean aggregate number of
transmitted messages in the network.} \label{fig.samplePath}
\end{figure}

The average message complexity of CRW can also be quantified by building on the asymptotic characterization of~\cite{Cox:1989}.
Towards that end consider a typical sample path of the process $(\sum_i\xi_i(t)\mid t > 0)$ illustrated in Figure~\ref{fig.samplePath}. By Lemma~\ref{lem:av_mes}  the average {\em aggregate} message complexity of the algorithm is the mean of the shaded area under the trajectory of $\sum_i\xi_i(t) : 0<t\le \tau_C$. The average per-node message complexity is then this quantity divided by the total number of nodes $n$.

Let
\[m_N=\left\{\begin{array}{ll}
N^{2} & \mbox{ if } d = 1 \\
N^2(\log N)^2 & \mbox{ if }d = 2 \\
N^d\log N & \mbox{ if }d \geq 3.
\end{array} \right.\] The following theorem provides an upper bound for
the message complexity of CRW.

\begin{theorem}\label{thm.msgComplexityC} \cite[Theorem 7]{SavasCISS}
\[\limsup_{N\rightarrow\infty}E[\eta_C(\tau_C)]/m_N~<~\infty.\]
\end{theorem}

To complement Theorem~\ref{thm.msgComplexityC} a lower on the growth rate of $E[\eta_C(\tau_C)]$ is provided by Theorem~\ref{thm.timeComplexityC}: Since $\sum_i\xi_i(t)\ge 1$ for all $t$  it follows that $E[\eta_C(\tau_C)]\ge E[\tau_C]$; in turn
\[\liminf_{N\rightarrow\infty}E[\eta_C(\tau_C)]/s_N~\ge~ \liminf_{N\rightarrow\infty}E[\tau_C]/s_N~>~0.\]
By comparing $s_N$ and $m_N$ we conclude that this bound is tight for a ring, i.e. a 1-$d$ torus, and it is off by at most a factor $\log N$ for $d \ge 2$. For a relatively concrete view of these quantities Figure~\ref{fig:torus} provides a summary of numerical simulations for time and message complexities of CRW on 2-dimensional tori of varying sizes.
%
%

\section{Two-Phase Algorithms} \label{sec.twophase}
The algorithm would work as follows: In the first phase the nodes in the network follow a CRW protocol upto some designated deterministic time $t$. At this time there are $\eta_C(t)$ tokens left in the system. The set of nodes that have tokens at time time $t$ then \emph{flood} their messages to all the nodes as described below.

Controlled Flooding (CFLD) algorithm assumes no network topology. It works by forwarding the message over all links. From the source node message is sent to all neighbors. Each node, $v$, receiving its first message from vertex $u$ sends messages to all neighbors except $u$. Also, each node will transmit its packet at most once~\cite{bertsekas}. If no message is received then it does nothing. It is well known that the message complexity scales as $\Theta(|edges|)$ since each edge delivers the message either once or twice~\cite{bertsekas}. The time complexity scales as $\Theta(diam(G))$ since we must reach all nodes~\cite{heinz}. We can adapt CFLD algorithm to our scenario where we have a random number $\eta_C(t)$ tokens left in the system. At time $t$ all nodes cease to implement CRW. Nodes with tokens separately CFLD messages. Each node then fuses the data once all the messages are received.

There are two fundamental questions that arise: \\(1) How do nodes realize that they have all the messages; \\ (2)Can we ensure that consensus is achieved through this process.\\ The answer to the second question lies in Lemma~\ref{lemma.invariant}, which asserts that the data fusion is invariant to order of reception. Consequently, we are left to  address the first requirement. Here we invoke the secondary distributed computation scheme described in Figure~\ref{fig.PCode} and Section~\ref{sec.algo}. Each active node at time $t$ has its individual variable {\tt count}. During the CFLD phase each node forwards this count variable in addition to its fused value. Each node can then determine whether it has received all the messages by updating its private {\tt count} variable.

There are three principal advantages for combining CRW and CFLD:\\
({\bf A}) Consensus is obtained in finite time.  \\
({\bf B}) CRW slows down when there are few tokens increasing time complexity. The two phase algorithm substantially improves time complexity. \\
({\bf C}) Analytical bounds for message and time complexities for general graphs can be established. This is because it is easier to determine the expected number of tokens left in the system at any time.


Next we will present message and time complexity for the combined algorithm. Recall the n-node communication graph $G=(V,E)$ with link set $E$ and nodes $V$; Let $d_v$ denote the degree of node $v\in V$; and $\mbox{diam}(G)$ denote the diameter of the graph.
To simplify the exposition we denote,
\begin{equation} \label{eq:tokens}
N(t) = E \left [\sum_{i=1}^n \xi_i(t) \right ]
\end{equation}
Note that $1 \leq N(t) \leq n$ and $N(0) = n$, and $\xi_i(t)$ is as before the state of node $i$ at time $t$.



We compute the time it takes for the expected number of active tokens to be below some positive integer $\gamma$. This leads to the following definition.
\begin{definition}
$\gamma$-time complexity is the time $\mathbf{T}_{\gamma}$ it takes for the CRW on an n-node graph to have an average of $\gamma$ active tokens left in the system, i.e.,
$$
\mathbf{T}_{\gamma}= \min \left \{t \geq 0 \mid N(t) \leq \gamma \right \},\,\, \gamma = 2,\,3,\,\ldots,\,n
$$
\end{definition}
Observe that unlike the termination time, $\tau_C$ defined in the Section~\ref{sec.torus}, $\mathbf{T}_{\gamma}$
is no longer a random variable, which as we will see in Section~\ref{sec.arbitrary} will simplify our analysis.

Define message complexity for the CRW until time $t$:
\begin{equation} \label{eq:gammsg}
M_C(t)=\int_0^t N(s) ds.
\end{equation}

The time complexity, $\mathbf{T}$, is the sum of the time complexities corresponding to the two phases. Consequently, for the case when all the tokens during the CFLD phase are transmitted at a unit rate we have:
\begin{equation} \label{eq.timecomp}
\mathbf{T} \leq \mathbf{T}_{\gamma} + O(\mbox{diam}(G))
\end{equation}

\begin{theorem} \label{thm:twophase}
The overall message complexity, $M(t)$ for the two phase scheme where nodes follow CRW upto time $t$ and CFLD after time $t$ is less than $M(t)+2(\sum_{v\in V} d_v)N(t)$. For $t = \mathbf{T}_{\gamma}$, the overall message complexity is less than $M_C(\mathbf{T}_{\gamma})+ 2(\sum_{v\in V} d_v) N(\mathbf{T}_{\gamma})$.
\end{theorem}
\begin{proof}
Suppose, the CFLD phase starts at time $t$ then the overall number of messages, $\eta(t)$ is the sum of messages transmitted during the CRW phase, $\eta_C(t)$, and that transmitted during the CFLD phase, $\eta_F(t)$ starting at time $t$. Specifically, we have $\eta(t) = \eta_C(t) + \eta_F(t)$. Taking expectations on both sides we obtain: $E(\eta(t)) = M(t) + E(\eta_F(t))$. We can simplify the expression for the second phase by noting that(see \cite{bertsekas}),
$E(\eta_F(t)) \leq 2(\sum_{v\in V} d_v) N(t)$. The proof now follows by substitution.
\end{proof}

\section{Time and Message Complexities for General Graphs}\label{sec.arbitrary}
This section describes techniques for estimating the message and time complexity of CRW for general graphs in both continuous and discrete time settings. Bounds for SRW reduces to computation of cover times. Cover time bounds for many of the graphs considered in this paper are available in the literature and we do not develop these results here.

To develop results for CRW we will follow the two-phase procedure outlined in the previous section. The advantages of the two-phase algorithm has already been outlined in Section~\ref{sec.twophase}. We recall one main advantage that is pertinent here, namely, from Theorem~\ref{thm:twophase} it follows that we do not have to seek bounds for the stopping times $\tau_C$. Rather we only need to determine the expected number of active tokens at a deterministic time $t$.

This section is organized as follows. First we establish straightforward results for general graphs based on bounds on the worst-case mean hitting time. We show that the number of active tokens at time $t$ decays as $O(n\exp(-{t \over \sigma}))$, where $\sigma$ is the worst-case mean hitting time on the graph. We compute $\sigma$ network-circuit resistance analogy. We then compute complexity bounds for a number of graphs such as expanders and meshes. While this bound is general, it turns out to be conservative in estimating the message complexity. The main reason is that the worst-case mean hitting time is generally large for many graphs and a local analysis is required. This motivates a careful study of message complexity based on local analysis of random walks. Specifically, we consider graphs with geometric structure. We show that for such graphs the message complexity scales as $O(\log^2(n))$ paralleling our results for the torus in Section~\ref{sec.toruscomp}.

As before let $X_t$ denote the state of a random walk on a graph $G=(V,E)$. For continuous time we consider unit rate random walks and assume that multiple random walks are independent. Analogously we also consider discrete time independent simple symmetric random walks and allow self-loops in the graphs. Our definitions and results typically apply to both continuous and discrete setups based on the so called jump-and-hold description\footnote{Specifically, as described in \cite{aldous:book} the continuous walk can be constructed by the two step procedure, namely, (1) Run a discrete time chain with the simple symmetric transition matrix; (2) Given the sequence of states, $v_j \in V,\,\,j=1,\,2,\,\ldots,m$ visited by the discrete time chain, the duration of time spent at each state, $v_m$ is a unit rate exponentially distributed random variable. This continuization is particularly useful since useful quantities such as mean hitting times etc. in the continuous case corresponds directly to the mean number of discrete time steps required in the discrete time chain.} and a continuization argument as described in \cite{aldous:book}. Nevertheless, wherever appropriate we will point out specifically whether our results apply to discrete or continuous time scenarios.

Analogous to Equation~\ref{eq:gammsg} for the continuous time setup, the message complexity in the discrete setup is given by,
\begin{equation} \label{eq:gammsgdisc}
M_C(t)=\sum_{s=0}^t N(s)
\end{equation}

We denote the first hitting time
of node $v$ by $T_v$, i.e., $T_v = \inf\{t \geq 0 \mid X_t=v\}$. We also denote by $T_{vw}$ the hitting time for a random walk starting at $v$ and hitting $w$ for the first time.

$$
T_{vw} = \inf\{t \geq 0 \mid X_0=v,\,\,X_t=w\}
$$

The worst-case hitting time is denoted as $\sigma$, i.e.,
$$
\sigma = \max_{v,w \in V} E(T_{vw})
$$

Let $C_{vw}$ be the first time that two independent unit rate continuous time random
walks, $X_t,\,Y_t$, on graph $G=(V,E)$ started at nodes $v$ and $w$ coalesce(meet), i.e.,
$$
C_{vw} = \inf\{t \geq 0 \mid X_t=Y_t,,\,X_0=v,\,Y_0=w\}
$$

The meeting times for two independent copies of random walks in continuous time is related to the worst-case hitting time. Specifically, Aldous \cite{aldous:book} (Proposition 5, Chap 14) uses Martingale arguments to show that,
\begin{equation} \label{eq:meethit}
\max_{v,w} E(C_{vw}) \leq \sigma
\end{equation}

Let $\alpha_s(A)$ denote the worst-case coalescing time probability
on a subset $A \subset V$, i.e.,
\begin{equation} \label{eq:meethitprob}
\alpha_s(A) = \min_{v,w \in A} Prob(C_{vw} \leq s)
\end{equation}

Note that by union bounding we obtain, $\alpha_s(A) \geq \alpha_s(V)$. Now through Markov inequality together with Equation~\ref{eq:meethit} we get a bound on the probability of meeting time, i.e.,
\begin{equation} \label{eq:meethitbnd}
\alpha_s(A) \geq \alpha_s(V) \geq 1 - {\sigma \over s}
\end{equation}

We next consider decomposition of the original graph into disjoint subgraphs and bound the total coalescing time by the union of the coalescing times for the subgraphs.
Let $\lfloor t \rfloor$ denote the greatest integer smaller than
$t$. Suppose $A_i, i=1,\,2,\,\ldots,\,m(t)$ be a partition
of the vertices of the graph and $\ca_t$ denotes the collection,
i.e., $$\bigcup \limits_{i=1}^{m(t)} A_i = V,\,\,\,A_i \bigcap A_j = \emptyset,\,\, i
\neq j,\,\,\ca_t=\{A_1,\,A_2,\,\ldots,A_{m(t)}\}$$
The worst-case coalescing time, $\alpha_s(\ca_t)$, over this sub-collection is defined by
\begin{equation} \label{eq:wcmeetprob}
\alpha_s(\ca_t) = \min_{1\leq j \leq m(t)} \alpha_s(A_j)
\end{equation}
\begin{theorem} \label{p.ubdd}
Consider the partition of the graph into subsets, $\{A_k\}$, as described above. Suppose $1\leq m(t) \leq {N(t)\over 2}$ i.e.,  the number of partitions is smaller than one half the expected number of active tokens at time $t$. It follows for both continuous and discrete time setups that,
\begin{equation} \label{e.finalcontrac1}
N(t+s) \leq N(t) \exp\left ( -{1 \over 2} \alpha_s(\ca_t)\right );\,\,\,\,0\leq s\leq t,\,\,\,\,N(t) \geq 2.
\end{equation}
Furthermore, suppose $t \leq r \leq r + s \leq 2 t$ and the number of partitions are chosen such that $1\leq m(t) \leq {N(t)\over 4}$ and $N(t) \leq 2N(2t)$, then it follows that,
\begin{equation} \label{e.finalcontrac}
N(2t) \leq N(t) \exp\left ( -\left \lfloor {t \over 2 s} \right \rfloor \alpha_s(\ca_t)\right );\,\,\,\,0\leq s\leq t,\,\,\,\,N(t) \geq 2.
\end{equation}
\end{theorem}
The proof of the theorem appears in the appendix and is based on the arguments presented in Cox~\cite{Cox:1989} for the torus. We exploit the salient steps there to extend it to general graphs.

Observe that if the coalescing time of two walks is a constant then the number of active tokens decreases exponentially fast. However, the meeting time can be large, namely, the probability that two walks meet in a short time can be very small.
Note that since $0\leq \alpha_s(\ca_t) \leq 1$ the right hand sides of Equation~\ref{e.finalcontrac1} is larger than $N(t)/\sqrt{e}$. Consequently, Theorem~\ref{p.ubdd} is not useful for large incremental times $s$. Therefore, this result will be used as an intermediate step in an iterative process over many increments to provide useful bounds.


We will now use Theorem~\ref{p.ubdd} to prove the $\gamma$ time and message complexities for arbitrary connected graphs. We have the following theorem.
\begin{theorem} \label{thm:wkbnd}
Consider the algorithm CRW on an arbitrary connected graph, $G=(V,E)$. The $\gamma$ time complexity for $\gamma \geq 2$ scales\footnote{The $O(\cdot)$ notation here and in the rest of this section for time and message complexity implies that the bound holds for sufficiently large time for a fixed n-node graph.} as $O(\sigma\log(n/\gamma))$. The $\gamma$ message complexity for $\gamma \geq 2$ scales as $O(n\sigma\log(n/\gamma))$.
\end{theorem}
\begin{proof}
In Theorem~\ref{p.ubdd} we choose a single partition, i.e., $\ca_t = \{V\}$. For this case we note that the worst-case meeting time $\alpha_s(\ca_t)=\alpha_s(V)$. Consequently, we can apply Markov inequality described by Equation~\ref{eq:meethitbnd} to obtain $\alpha_s(V) \geq 1 - \sigma/s$. This bound only makes sense if $s > \sigma$. We choose time increments $s = 2\sigma$ and partition $T=4\sigma \log(n/\gamma)$ into $2\log(n/\gamma)$ increments. For each increment $s$ we obtain from Equation~\ref{e.finalcontrac} that,
$$
N(t+s) \leq N(t) \exp(-(1-\sigma/s)) = N(t) \exp(-1/2)
$$
Repeating this $2\log(n/\gamma)$ times we get
$$
N(T) \leq N(0) (\exp(-1/2))^{2\log(n/\gamma)} = \gamma
$$
where the last equality follows from the fact that $N(0)=n$.
The $\gamma$ message complexity directly follows from Equation~\ref{eq:gammsg}.
\end{proof}


In the next section we will now apply Theorem~\ref{thm:wkbnd} for specific graphs to obtain bounds on time and message complexities.
\subsection{Time and Message Complexity Based on Hitting Time Characterization}
Our goal in this section is to use well known bounds on hitting times for some well known graphs together with Theorem~\ref{thm:wkbnd}.

For general graphs Aleliunas et al~\cite{aleliunas} showed a general upper bound $\sigma = O(|E| |V|)$, for the worst-case hitting time, where $|E|$ is the number of edges and $|V|$ is the number of nodes (vertices). If the maximal degree of the graph is $D_{\max}$ then $|E| \leq n D_{\max}$ and $|V| =n$. This implies that the $\gamma$ time complexity scales as
$$\mathbf{T}_{\gamma}=O \left (n^2D_{\max}\log(n/\gamma)\right )$$
We note that this result is generally conservative in comparison to the time complexity of 2D torus described in the previous sections. This is because this hitting time bound is conservative. We invoke resistance characterization of hitting time to obtain sharper bounds.

\subsubsection{Resistance characterization for Connected graphs} \label{sec.resist}
 Chandra et al\cite{tiwari} establish bounds for hitting time between any two nodes based on resistance of electrical networks. Note that the resistance bounds apply generally to discrete time walks. However, note that there is a close relationship between the discrete and continuous time random walks based on the so called jump-and-hold description described earlier, which results in similar results for continuous time with appropriate time scaling.

 The electrical network is obtained by replacing each edge in the graph with a one-ohm resistor. It turns out that the worst-case mean hitting time satisfies
 \begin{equation} \label{e.resist}
 \sigma \leq \max_{u,v \in V}2|E|\rho_{uv} \stackrel{\Delta}{=} \rho^*
 \end{equation} where $\rho_{uv}$ is the effective resistance between nodes $u$ and $v$. Consequently, if $D_{\max}$ is the maximum degree for the graph and $\rho^*$ is the maximum effective resistance between any two nodes in the network, we get
\begin{equation} \label{e.msgresist}
\mathbf{T}_{\gamma} \leq 2 n \log(n/\gamma) D_{\max} \rho^*
\end{equation}
{\bf Expander Graphs:} An $(n,D_{\max},\alpha)$ expander is a graph $G=(V,E)$ on n vertices
of maximal degree $D_{\max}$ such that every subset $A\subset V$ satisfying $|A|\leq n/2$ has
$|N(A)- A| \geq \alpha |A|$, where
\begin{equation} \label{eq:gneib}
N(A) = \{v \in V \mid (u,v) \in E,\, u \in A\}
\end{equation}
For an $(n,D_{\max},\alpha)$ expander graph with minimum degree $D_{\min}$, the worst-case resistance is equal to $$\rho^* = {24 \over \alpha^2 (D_{\min} +1)}.$$ Consequently, the $\gamma$ time complexity scales as:
$$
\mathbf{T}_{\gamma} \leq 48 {n \log(n/\gamma) D_{\max} \over \alpha^2 (D_{\min} +1)},\, \gamma \geq 2
$$
For an expander graph, where $D_{\min} \approx D_{\max}$ we get a $\gamma$ time complexity scaling as $$\mathbf{T}_{\gamma} = O(n \log(n/\gamma))$$ We note that this result is close to the time complexity bounds obtained for a completely connected network in the previous section.

\noindent
{\bf 2D Mesh:} From the resistance calculations it turns out that $$\rho^* = O(\log(n))$$ for 2D mesh \cite{tiwari} with $n$ nodes. Consequently, for the 2D mesh we obtain
$$\mathbf{T}_{\gamma} = O({n \log^2(n/\gamma)}),\, \gamma \geq 2$$
This is within a $\log(n)$ factor of the bound obtained for the 2D torus using more elaborate martingale calculations in the previous section. Note that unlike the 2D torus the 2D mesh is not symmetric and results of the previous section cannot be directly applied here.

\noindent
{\bf Random Geometric Graphs (RGG):}
A 2D Random Geometric Graph with $n$ nodes and
radius $r(n)$, denoted by $G_{r(n)}=(V,E)$, is a
graph where nodes are uniformly distributed in the unit square and $(u,v) \in
E$ if and only if the Euclidian distance between nodes $u$ and
$v$ is smaller than or equal to $r(n)$.

It is well known that when the radius of connectivity is chosen as $r(n)=\sqrt{2 \log
n / n}$, the graph is connected with high probability. Furthermore, Avin and Ercal~\cite{avin04} (Theorem 5.3) show that, with high probability, the resistance scales as $$\rho^*=O(1/nr^2(n))$$ and the number of edges scales as $|E|=O(n^2r^2(n))$ (see \cite{avin04} Corollary 3.5) for this choice of connectivity radius. Consequently, the worst-case mean hitting time scales as $\sigma=O(n)$ with high probability. This implies that for geometric random graphs with $r(n) = \sqrt{2 \log
n / n}$ the $\gamma$ time complexity scales as $$\mathbf{T}_{\gamma} = O(n \log(n/\gamma)),\, \gamma \geq 2.$$

While the time-complexity bounds obtained using resistance characterization appears to be tight for several cases, the $\gamma$-message complexity is overly conservative. This is because the worst-case mean hitting time, $\sigma$ is $\Omega(n)$ in general. Theorem~\ref{thm:wkbnd} implies that the $\gamma$ message complexity scales as $O(n^2\log(n))$ even for a 2D torus. This is significantly weaker than the complexity bounds obtained for the torus in Section~\ref{sec.torus}. Motivated by these reasons we develop a new characterization of message and time complexities based on local geometric analysis of random walks.

\subsection{Logarithmic Bounds for Message Complexity}
The main conservatism in Theorem~\ref{thm:wkbnd} arises from the fact that the meeting time is bounded in terms of the worst-case hitting time. Specifically, if two random walks start relatively close to each other we expect that the meeting time is relatively small, i.e., the meeting time should typically scale with initial distance between the two walks. In this section we develop these ideas further for graphs that have a geometric neighborhood structure. We focus on discrete time walks since the analysis is technically simpler. Each active token follows an independent, simple, symmetric random walks on the graph $G=(V,E)$. Specifically, at each step an active token moves to a neighbor of its current location, chosen uniformly at random and the moves of all the active tokens are synchronized (this assumption is not restrictive since we allow self-loops).

The basic idea is based on local behavior of random walks. Specifically, it turns out that for graphs that are endowed with a geometric neighborhood structure it is possible to characterize the probability that two random walks meet in terms of their initial graph distance. We emphasize that while in general there is always a non-zero probability that two random walks meet, this probability has often been characterized in terms of the entire graph. Indeed this was the basic reason for the conservatism of resistance based bounds derived in the previous section. Therefore, to overcome this issue we will develop results based on local behavior of random walks.
Our main result in this section (see Theorem~\ref{thm:mainresult}) will establish that under certain regularity conditions on the graph the expected number of active tokens at time step $t$ decays inversely with $t$, i.e.,
\begin{equation} \label{e.scaling}
N(t) = O\left ({n\log(t+1)\over t}\right ),\,\, N(t) \geq 2
\end{equation}
\emph{We again emphasize that the $O(\cdot)$ notation above and in the rest of this section refers to time asymptotics for a fixed n-node graph}.
The result implies a bound on both $\gamma$ time complexity and $\gamma$ message complexity. The $\gamma$ time complexity scales as
$$
\mathbf{T}_{\gamma} \leq C {n \log(n) \over \gamma},\,\, \gamma = 2,\,3,\ldots, n
$$
Note that the $\gamma$ time complexity bounds are order-wise similar to those derived using resistance arguments in the previous section. However, the main advantage here is that we can now obtain a bound on message complexity based on Equations~\ref{eq:gammsgdisc},~\ref{eq:strngbnd}:
\begin{equation} \label{eq:strngbndmsg}
M(\mathbf{T}_{\gamma}) \leq C \sum_{t=1}^{\mathbf{T}_{\gamma}} {n \log(1+t) \over \gamma t} \leq C {n\over \gamma}\left (\log^2(n) + o(\log(n))\right )
\end{equation}
Thus the message complexity per node scales as $n^{-1} \eta_{\gamma} = O(\log^2(n))$.

This result is based on the fact that for many graphs the probability that two walks at a distance $R$ meet in time $R^2$ is bounded from below by the $1/\log(R)$. To precisely describe these ideas we introduce some notation. Let $d(u,v)$ be the graph distance between the nodes $u,\,v\in V$, i.e., the minimal number of edges in any edge path connecting $u$ and $v$. We denote by $B(u,R)$ the ball centered at node $u$ and radius $R$, i.e.,
$$
B(u,R) = \{v \in V \mid d(u,v)<R\}
$$
The volume of a set, $A \subset V$, denoted by $Vol(A)$, is the number of edges contained in the ball. The volume of the ball, $B(u,R)$ is denoted by $Vol(u,R)$ for simplicity. Note that if $d_v$ is the degree of node $v$ then we have,
$$
Vol(u,R) = Vol(B(u,R)) = \sum_{v \in B(u,R)} d_v
$$

Next we denote by $P(u,v)$ the 1-step transition probability of going from node $u$ to node $v$. Since we consider simple symmetric random walks, this transition probability is the inverse of the degree of node $u$ if $u$ and $v$ are connected and zero otherwise. We also use $P_t(u,v)$ to denote the t-step transition probability for going from $u$ to $v$. We next present a precise characterization when Equation~\ref{e.scaling} holds. We will see that this bound holds when one has a geometric neighborhood structure as described below:

\begin{definition}
A Graph $G=(V,E)$ is said to satisfy a geometric neighborhood structure if there exists constants, $C_0,\,C_1$ such that
\begin{align} \label{e.quadgrowth}
C_0 R^2 \leq |B(u,R)|; \,\,\, |B(u,R) \bigcap B(u,R+\Delta)| \leq C_1\Delta R ,\,\,\, \forall\,\, u \in V,\,\,0< \Delta \leq R.
\end{align}
where, $0\leq R \leq R_{\max}$ and $R_{\max}$ is the diameter of the graph.
\end{definition}

Typically a graph that is approximately regular and has a geometric neighborhood structure satisfies such a property. The geometric random graph described earlier asymptotically satisfies the geometric neighborhood property. Indeed, note that due to the uniform distribution of the nodes in the unit cube this property is satisfied for sufficiently large $n$ with high probability (see \cite{avin04} for more details). The theorem below will evidently require only the lower bound. However, it turns out that to ensure a suitable bound on the meeting time probability the upperbound will also be necessary.

\begin{theorem} \label{thm:mainresult}
Suppose the graph $G=(V,E)$ has a geometric neighborhood structure as described in Equation~\ref{e.quadgrowth} and the meeting time probability satisfies:
\begin{equation} \label{eq:meetlbnd}
\alpha_{t^2}(B(u,t)) \geq {C_2 \over \log(t)},\,\,t > 0,\,\,u \in V
\end{equation}
for some constant $C_2$ independent of time $t$.  Note that $\alpha_{t^2}(B(u,t))$ is the meeting time probability (see Equation~\ref{eq:meethitprob}) for any two walks starting in the ball $B(u,t)$ in time $t^2$.
Then the expected number of active tokens at time step $t$ satisfies
\begin{equation} \label{eq:strngbnd}
N(t) \leq C \left ({n\log(t+1)\over t}\right ),\,\,t \geq 1
\end{equation}
where $C = {16 \over C_0} \max( 8 , {\log(2) \over C_2})$ when the number of active tokens is greater than $4$.
\end{theorem}
Note that smaller the constant $C_0$ the larger the number of active tokens at time $t$.
We are now left to determine the conditions under which Equation~\ref{eq:meetlbnd} is satisfied.
Surprisingly, it turns out that the logarithmic bound holds if:\\ (a) The t-step transition probability is approximately Gaussian. \\ (b) Geometric neighborhood property as described in Theorem~\ref{thm:mainresult} holds. \\
This result is stated below.

\begin{lemma} \label{lem:tsteptrans}
Consider the graph $G=(V,E)$ satisfying the geometric neighborhood property as described in Equation~\ref{e.quadgrowth}.  Suppose the t-step transition probability satisfies the so called Gaussian bound, i.e.,
\begin{equation} \label{eq:tsteptrans}
{C_3 \over t} \exp\left (-{d^2(u,v) \over C_4 t} \right ) \leq P_t(u,v) + P_{t+1}(u,v);\,\forall \,\,u,\,v \in V,\,\, 1 \leq d(u,v) \leq t
\end{equation}
where $C_3,\,C_4,$ are positive constants independent of time. Then the probability of meeting time satisfies Equation~\ref{eq:meetlbnd} for some suitable constant $C_2$. Consequently, these conditions also imply the $\gamma$ message complexity bound described by Equation~\ref{eq:strngbnd}.
\end{lemma}
Note that the Gaussian t-step transition estimate bounds the sum of the transitions at $t$ and $t+1$. Note that for bi-partite graphs we must have either $P_t$ or $P_{t+1}$ equal to zero.  Therefore, we cannot hope to improve this situation in general. However, if each node has self-loops it turns out that we can lower bound the $t$ step transition probability directly, i.e., for non-bipartite graphs we have
\begin{equation} \label{eq:tsteptrans1}
{C_3 \over t} \exp\left (-{d^2(u,v) \over C_4 t} \right ) \leq P_t(u,v);\,\forall \,\,u,\,v \in V,\,\, 1 \leq d(u,v) \leq t
\end{equation}

Our problem now reduces to finding those graphs that satisfy the t-step Gaussian transition property. It turns out that weak homogeneity conditions lead to the Gaussian t-step transition property. We describe what these conditions are next.

\paragraph{Volume Doubling Property:}
A graph $G=(V,E)$ is said to satisfy \emph{volume doubling property} if volume of a ball centered at any point, $u$, with increasing radius satisfies
\begin{equation} \label{eq:voldoub}
Vol(u,2R) \leq C_5 Vol(u, R),\,\, \forall \,\, u \in V,\,\, R>0
\end{equation}

We again point out that a 2D mesh satisfies such a property. The volume at a graph radius $R$ and $2R$ is smaller than $R^2$ and $4R^2$ respectively for any $R$. A similar result holds for random geometric graphs(RGG) due to the so called geo-dense property \cite{avin04}. For a constant $\mu\geq 1$, a graph is said to be $\mu$-geo-dense
if every square bin of size $A\geq r^2(n)/\mu$  (in the unit square) has $nA$ nodes. Recall from Section~\ref{sec.resist} that any two nodes at a Euclidean distance $r(n)$ is connected. Lemma 3.2 of \cite{avin04} shows that with high probability if $r^2(n) = c\mu \log(n)/n$ then RGG is $\mu$ geo-dense. Furthermore, if RGG is $\mu$ geo-dense then, (i) Each node, $v$, has degree $d_v = \Theta(nr^2(n))$; (ii) $|E| = \Theta(n^2r^2(n))$. Consequently, we immediately see that the volume doubling property holds since RGG is evidently close to a 2D mesh in terms of volumes at the different radii except for a $\log(n)$ factor.

\paragraph{Constant Resistance Property:} For any subsets $A \subset B \subset V$ consider an electrical network with one-ohm resistors for each edge on the graph $G=(V,E)$. Define the resistance, $\rho(A,B)$, between $A$ and $B$ as the power dissipated when a one-volt potential is applied to all the nodes in $A$ and the nodes in the complement of $B$, i.e., $B^c$ are all grounded. The graph $G=(V,E)$ is said to satisfy the constant resistance property if:
\begin{equation} \label{eq:resist}
C_6 \leq \rho(B(u,R), B(u,MR))  \leq C_7
\end{equation}
where $M$ is any number larger than one and $C_6$ and $C_7$ are constants that can depend on $M$ but not on $R$.

Again consider first the example of a 2D mesh. Due to symmetry all the nodes at distance $R+\Delta$ have the same potential. Consequently we can short all the nodes at this distance. Due to the geometric neighborhood property, there are about $R+\Delta$ nodes connected to nodes at a distance $R+\Delta-1$. Since this is a parallel set of resistances the effective resistance is $1/(R+\Delta)$. Summing over these resistances we obtain,
$$
\rho(B(u,R), B(u,MR)) = \sum_{\Delta=1}^{MR} {1 \over R+\Delta} \approx \log({MR \over R}) = \log(M)
$$
which establishes the fact. A similar but more elaborate argument is required for RGG. Basically the short cut principle along with the geo-dense property ensures a lower bound of $\Omega(\log(M))$. To obtain an upper bound we need to construct a flow along the lines of \cite{avin04} that satisfies the Kirchoff current law.

\paragraph{Uniform Isoperimetry Property:} We consider the subgraph, $G(u,R)$, formed by restricting the graph $G=(V,E)$ to the subset of vertices in the ball, $B(u,R)$. Consider any partition of $G(u,R)$ into $S,\,S^c$. We say that the graph $G=(V,E)$ satisfies a uniform isoperimetry property if for every $u$ and every $R$ we have,
\begin{equation} \label{eq:isoperimetry}
{C_8 \over R} \leq {Cut(S,S^c) \over \min(Vol(S),Vol(S^c))}
\end{equation}
where $C_8$ is some constant independent of $R$ and $Cut(S,S^c) = \sum_{u \in S,\,v \in S^c} \bf{1}_{u,v}$.

For the 2D mesh this is a well-known property (see \cite{chung}). The corresponding property for an RGG is a direct consequence of Theorem 4.1 of \cite{avin04}.

We are ready to state our result.
\begin{lemma}
Consider a graph $G=(V,E)$ that is in general infinite and satisfies the properties described in Equations~\ref{eq:voldoub},~\ref{eq:resist},~\ref{eq:isoperimetry}, then the t-step transition probability satisfies the Gaussian estimate described in Equation~\ref{eq:tsteptrans}.
\end{lemma}
\begin{proof}
The proof is a direct consequence of the results in Merkov \cite{mathussr86} and Grigoryan and Telcs \cite{telcs}. Theorem 3.1 in Grigoryan and Telcs \cite{telcs} states that if a graph $G=(V,E)$ satisfies the volume doubling property, the resistance property and the Elliptic Harnack Inequality, the t-step transition matrix satisfies Equation~\ref{eq:tsteptrans}. Merkov \cite{mathussr86} shows that the isoperimetry property implies the Elliptic Harnack inequality.
\end{proof}

%

\subsection{Message and Time Complexity for Achieving Consensus} \label{sec.mtc}
We will utilize Theorem~\ref{thm:twophase} to characterize message and time complexity for achieving consensus in general graphs. For general graphs we note from the resistance arguments of Equation~\ref{e.msgresist} that,
$$
\mathbf{T}_{\gamma} \leq 2 n \log(n/\gamma) D_{\max} \rho^* \implies \mathbf{T} \leq \mathbf{T}_{\gamma} + O(diam(G)) \leq 2 n \log(n/\gamma) D_{\max} \rho^* + O(n)
$$
As we described earlier this bound is not useful for characterizing message complexity. To obtain better bounds we restrict our attention to graphs satisfying volume doubling, constant resistance and uniform isometry described in the previous section. We note that the message complexity from Theorem~\ref{thm:twophase} can be bounded as:
$$
M(\mathbf{T}_{\gamma}) \leq M_C(\mathbf{T}_{\gamma}) + 2 \gamma \sum_{v \in V} d_v \leq O( {n\over \gamma} \log^2(n)) + 2 \gamma \sum_{v \in V} d_v
$$
where, $d_v$ is the degree of node $v$ and we have used Equation~\ref{eq:strngbndmsg} to determine a bound on $M_C(\mathbf{T}_{\gamma})$. We now let $\gamma = \log(n)$. It follows that the message complexity for the two phase scheme is:
$$
M(\mathbf{T}_{\gamma}) \leq O( n \log(n)) + 2 \gamma \sum_{v \in V} d_v \implies M(\mathbf{T}_{\gamma}) \leq O(n\log(n))
$$
where in the final inequality we have used the fact that $d_v \leq 4$ for two-dimensional Grid graphs. The time complexity for grid graphs follows from Equation~\ref{eq.timecomp},
$$
\mathbf{T} \leq \mathbf{T}_{\gamma} + O(\mbox{diam}(G)) \leq O(n)
$$
where we note that $\mathbf{T}_{\gamma}$ for $\gamma=O(\log(n))$ scales as $O(n)$ and $O(\mbox{diam}(G))$ scales as $O(\sqrt{n})$.

For RGG we note that with high probability the number of links is of $O(\log(n))$. Consequently, following along the same lines as the previous computation for 2D grid graphs we obtain
$$
M(\mathbf{T}_{\gamma}) \leq O(n\log^2(n))
$$
By noting that for RGG $\mbox{diam}(G) = O(\sqrt{n/\log(n)})$ we get a bound on the time complexity:
$$
\mathbf{T} \leq O(n \log(n/\gamma)) + O(\mbox{diam}(G)) \leq O(n)
$$
\subsection{Numerical Results}
\begin{figure*}\vspace*{-0.5cm}
\centerline{\subfigure[Per-node message
complexities]{\includegraphics[width=3.1in]{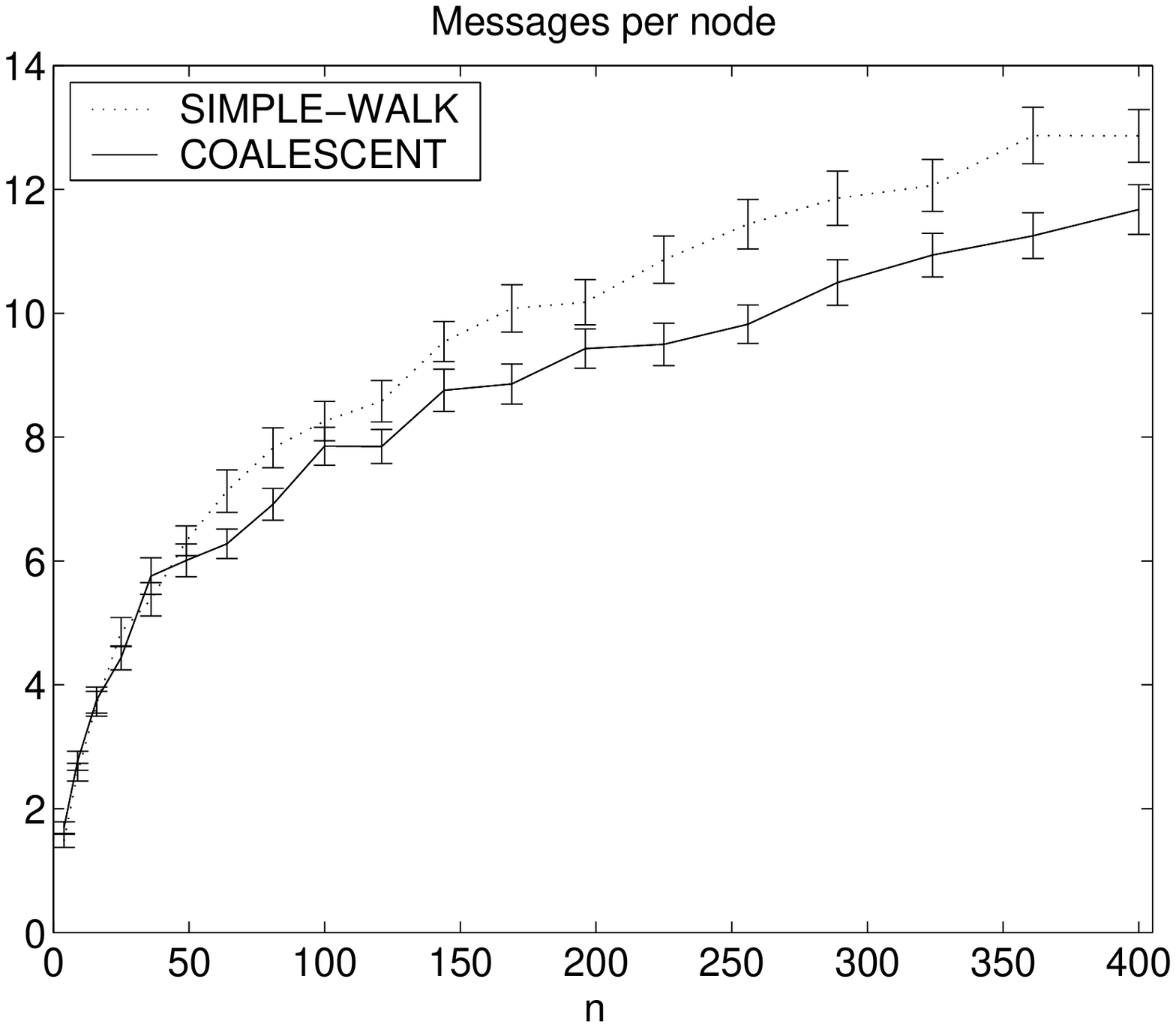}}
\hfil \subfigure[Time
complexities]{\includegraphics[width=3.2in]{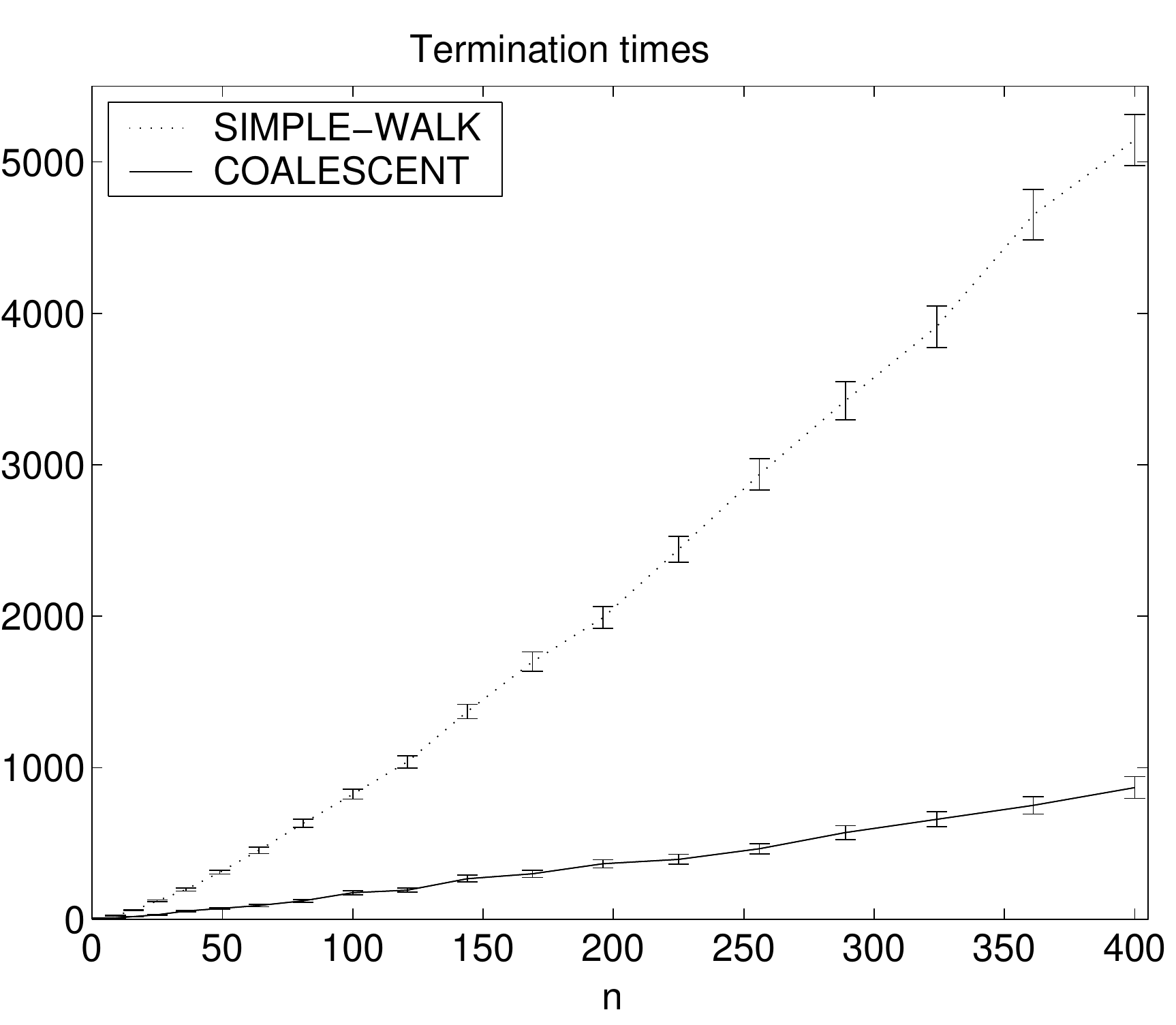}}}
\caption{\small Average execution times and message complexities per node for CRW and GOSSIP on the
2-dimensional torus with $n$ nodes. Note for an accurate comparison time and message complexity of CFLD needs to be added to that of CRW. Nevertheless, both time and message complexity are insignificant relative to that of CRW (see Section~\ref{sec.mtc}) for the torus.} \label{fig:torus}
\end{figure*}

Numerical verification of the analytical results of
Table~\ref{table.complexity} on a $2-d$ torus is presented in Fig.~\ref{fig:torus}.
Figure~\ref{fig:torus} provides a summary of numerical simulations for time and message complexities of CRW on 2-dimensional tori of varying sizes.

An important consideration is that GOSSIP achieves consensus at all nodes while SRW and CRW realize their solution at a random node. Therefore, strictly speaking for the comparisons to be meaningful we need to add the time and message complexities to obtain similar consensus estimates for SRW and CRW. We can obtained consensus through CFLD. The time complexity of CFLD for torii scales as $O(\sqrt{n})$, which is insignificant relative to time complexity of SRW/CRW. Message complexity-per-node of CFLD on torii scales as $O(\log(n))$, which is again insignificant relative to message complexity of CRW $O(\log^2(n))$. Consequently, the qualitative nature of the plots is similar even when we incorporate these additional costs.

We also simulate numerically time and message complexity for random geometric graphs. To simulate a 2 dimensional geometric random graph we distributed $n$ nodes in a unit square and formed edges whenever two nodes were at a distance smaller than $\sqrt{2
\log n / n}$. We discarded graphs that were not connected. Again to compare CRW/SRW against GOSSIP consensus costs must be incorporated. We can obtained consensus through CFLD. The time complexity of CFLD for RGG scales as $O(\sqrt{n/\log(n)})$, which is insignificant relative to time complexity of SRW/CRW. Message complexity-per-node of CFLD on RGG scales as $O(\log(n))$, which is again insignificant relative to message complexity of CRW $O(\log^2(n))$. Consequently, the qualitative nature of the plots is similar even when we incorporate these additional costs.

We next describe Gossip algorithm as studied in ~\cite{Boyd:2005} for the sake of completion. Gossip algorithms refer to distributed randomized algorithms that
are based on pairwise relaxations between randomly chosen node
pairs. In the present context a pairwise relaxation refers to
averaging of two values available at distinct nodes. In what follows a stochastic matrix
$P=[P_{ij}]_{n\times n}$ is called {\em admissible} for $G$ if
$P_{ij} = 0$ unless nodes $i$ and $j$ are neighbors in $G$. The
algorithm is parameterized by such a $P$:

{\bf Algorithm} {\tt GOSSIP-AVE}$(P)$: Each node $i$ maintains a
real valued variable with initial value $z_i(0)=x_i$. At the tick
of a local Poisson clock, say at time $t_o$, node $i$ chooses a
neighbor $j$ with respect to the distribution
$(P_{ij}:j=1,2,\cdots,n)$ and both nodes update their internal
variables as $z_i(t_o)=z_j(t_o)=(z_i(t_o^-)+z_j(t_o^-))/2$.
We associate each node with a real value and consider the problem of computing its mean value.
In order to make a fair comparison of \texttt{GOSSIP-AVE} with
\texttt{CRW} and \texttt{SRW} we need to use a stopping criterion for \texttt{GOSSIP-AVE}.
Let $\bar{x}$ denote the average of $x_1,x_2,\cdots,x_{n}$, let
$z(t)$ denote the vector $(z_1(t),z_2(t),\cdots,z_{n}(t))$ of node
values at time $t$, and $\mathbf{1}$ denote the vector of all 1s.
Define $\tau_k$ as the $k$th time instant such that some local
clock ticks and thereby triggers messaging in the network. For
$\varepsilon>0$ let the deterministic quantity $K(\varepsilon,P)$
be defined by
\[
K(\varepsilon,P)=\sup_{z(0)} \inf \left\{ {k \mbox{ }:\mbox{
}Pr\left( {\frac{{{\|z(\tau_k) - \bar{x}\mathbf{1}\|}_2}}
{{{\|z(0)\|}_2}} \geqslant \varepsilon } \right) \leqslant
\varepsilon } \right\}.
\]
In~\cite{Boyd:2005} $K(\varepsilon,P)$ is considered as a
termination time for Algorithm {\tt GOSSIP-AVE}$(P)$ and
minimization of $K(\varepsilon,P)$ is sought by proper choice of
$P$. Here we adopt the same interpretation for comparison
purposes. It should perhaps be noted here that this is a fairly
weak stopping criterion as
$\|z(\tau_{K(\varepsilon,P)})-\bar{x}\mathbf{1})\|_\infty/|\bar{x}|$
may be much larger than $\varepsilon$.


\begin{figure}
\centerline{\subfigure[Per-node message
complexities]{\includegraphics[height=2.5in]{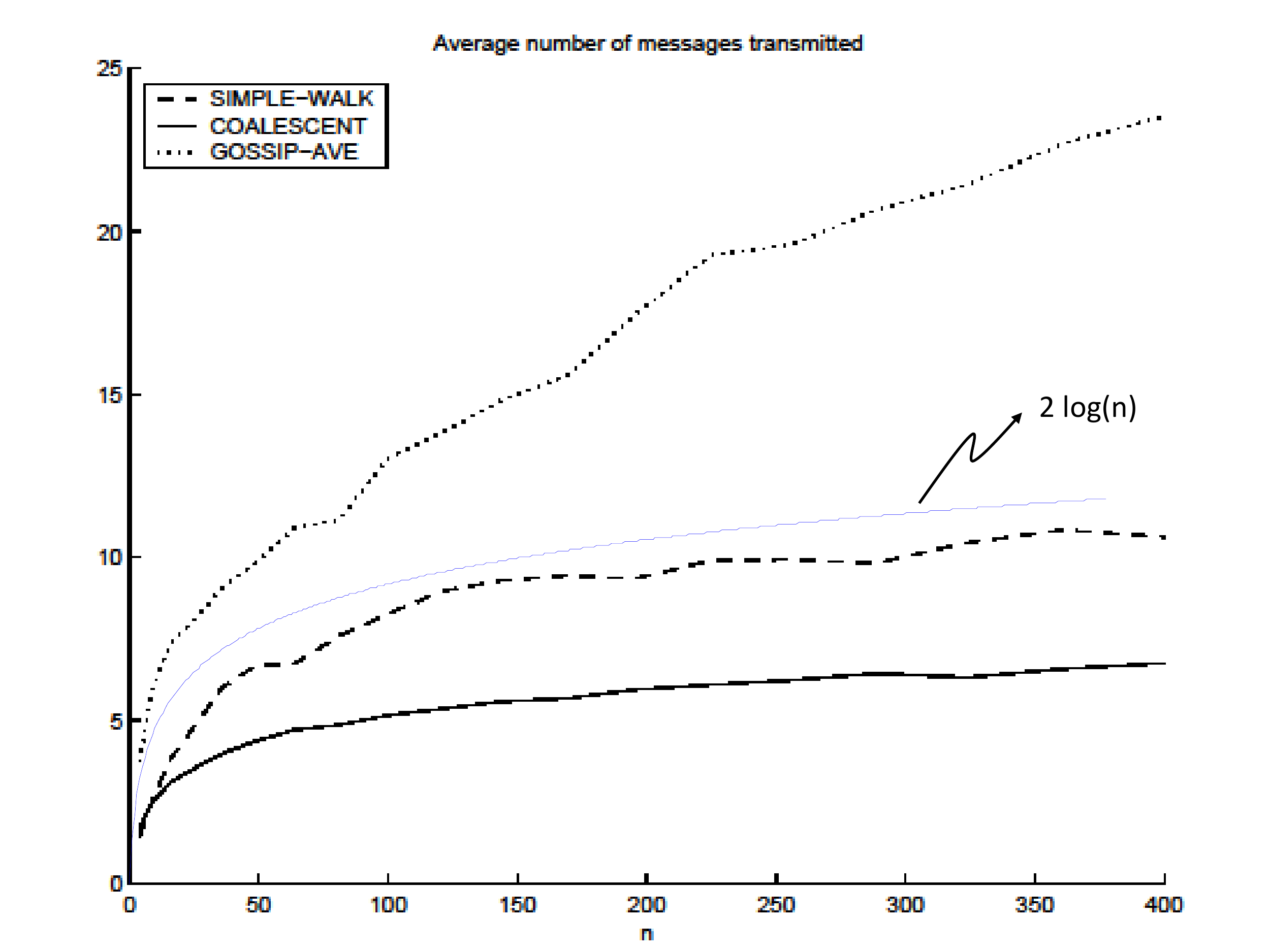}}
\hfil \subfigure[Time
complexities]{\includegraphics[height=2.5in]{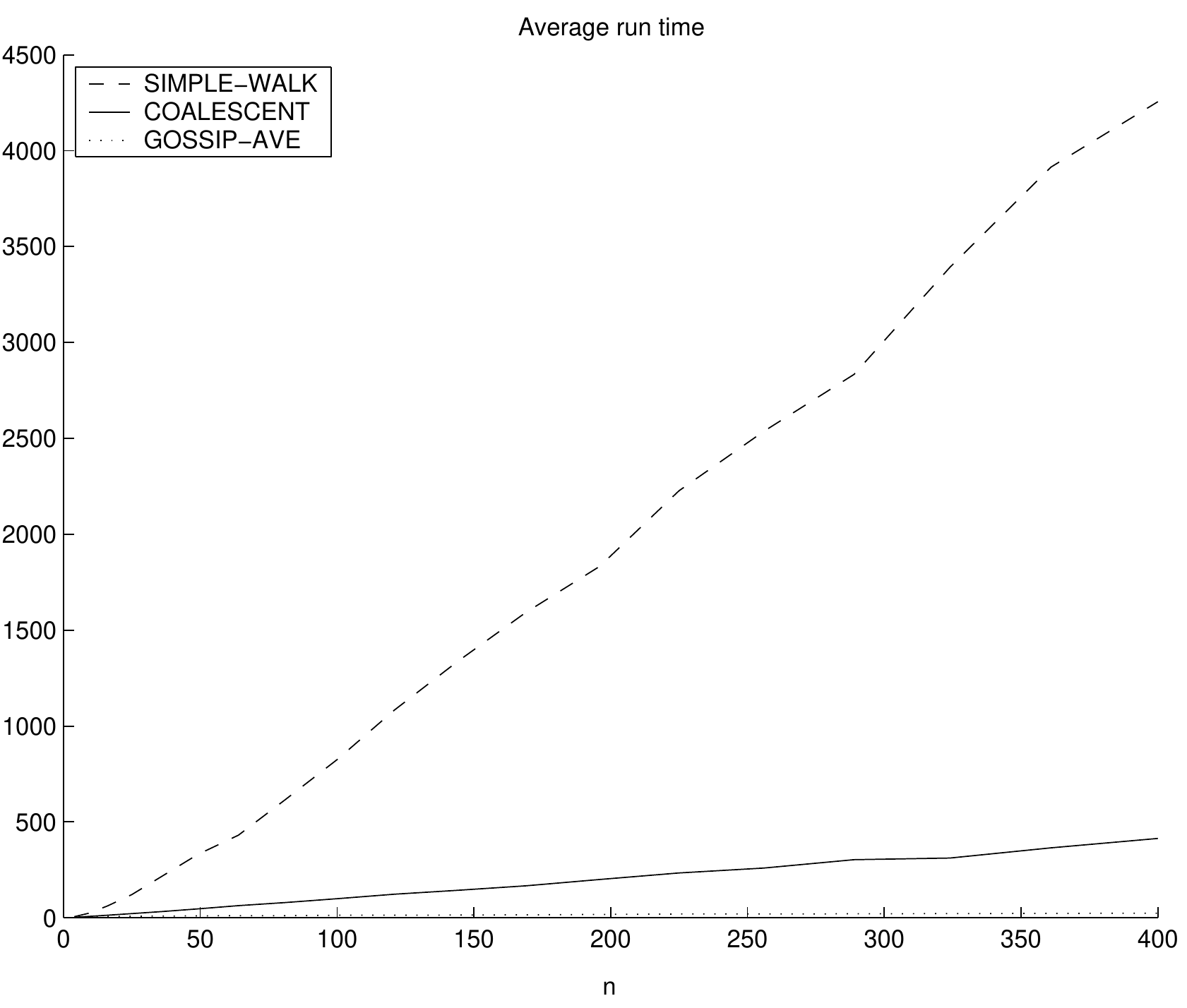}}}
\caption{\small Average run times and number of messages per node on
random geometric graphs (RGG). The solid curve represents
simulation results for {\tt CRW}, the dashed curve for
\texttt{SRW} whereas the dotted curve represent lower
bounds for {\tt GOSSIP-AVE}$(P)$ based on a lower bound for
$K(\varepsilon,P)$. Note also that exact value of
$F_n(x_1,x_2,\cdots,x_n)$ is obtained at the
termination of {\tt CRW} or \texttt{SRW} whereas no
such claim can be made for {\tt GOSSIP-AVE}$(P)$. Note for an accurate comparison, time and message complexity of CFLD needs to be added to that of CRW. Nevertheless, for RGG both time and message complexity are insignificant relative to that of CRW (see Section~\ref{sec.mtc}).}
\label{fig:geo}
\end{figure}

The numerical results of average number of messages and run times
for 2-D torus appear in Figure~\ref{fig:torus}. The corresponding results for geometric random graphs and a illustrative comparison with Gossip is presented in Fig.~\ref{fig:geo}. We have also plotted a bound $2\log(n)$ for comparison purposes. Notice that from the scale of the two plots it should be clear that the bound will have a similar qualitative relationship to the message complexity for the torus. These bounds reveal that the empirical per node message complexity appears to be closer to $O(\log(n))$ which is much smaller than the $O(\log^2(n))$ theoretical message complexity bound of Equation~\ref{eq:strngbndmsg}. One possibility for this difference is that our theoretical message complexity bound is for worst-case distribution of initial node values, while the empirical result is for an average case distribution of the node values.

\section{Appendix}

\subsection{Proof of Theorem~\ref{p.ubdd}}
We follow the argument of \cite{Cox:1989} and provide a detailed proof to point out that the proof goes through for general graphs. The proof applies to both discrete and continuous settings and basically utilizes Markovianity.
Let $\Lambda_B(t): t \geq 0,\, B \subset V$ denote the occupied
nodes (state) at time $t$ of a coalescing random walk whose initial
state is $B$. Observe that irrespective of the initial state, $B$,
if $0\leq s \leq t$, $E(|\Lambda_B(t)|) \leq E(|\Lambda_B(s)|)$.

We then have the following lemma:
\begin{lemma} \label{lem:car_bound}
Suppose $(X_t\mid t \geq 0)$ is a simple symmetric unit rate continuous
time random walk on graph $\Gamma$ and $B \subset A \subset V$, then
\begin{equation}
E[|\Lambda_B(s)|] \leq |B| -(|B|-1)\alpha_s(A). \label{eq:001}
\end{equation}
\end{lemma}

\begin{proof}
If $B = \emptyset$, Equation~(\ref{eq:001}) trivially holds
therefore assume $B$ is non-empty. Our approach is to find an upper
bound for the number of coalescences occurring in the time interval
$[0,s]$. Our analysis begins with starting random walks with
active tokens at $v \in B \backslash w$ and at $w$, and see whether
their paths ever meet. To formalize this approach define an
indicator function ${\cal I}(\cdot)$ to indicate whether or not
active tokens $v$ and $w$ meet in time $t$. Then,
\begin{equation} \label{e.conser}
Z(s) = \sum \limits_{v \in B \backslash w} {\cal I}(C_{v,w} \leq s)
\end{equation}
the random quantity $Z(s)$ is the number of active tokens $B
\backslash w$ which coalesces with $w$ at some time $0 \leq t
\leq s$. It follows through conservation of active tokens that,
$$
|\Lambda_B(s)| \leq |B| - Z(s)
$$
Now, $$E(Z(s)) = \sum_{v \not= w} Prob(C_{vw}\leq s) \geq \min_{v,w
\in B}Prob(C_{vw}\leq s)(|B|-1) = \alpha_s(A)(|B|-1)$$ The result now
follows by taking expectations on both sides in
Equation~\ref{e.conser} and substituting the above expression.
\end{proof}

Now consider the partition $A_i, i=1,\,2,\,\ldots,\,m(t)$ of the vertices of the graph as in the hypothesis of the theorem and let $B_j = A_j \cap B$.
\begin{lemma} \label{lem.icoal}
$|\Lambda_B(s)| \leq \sum_{j=1}^{m(t)} |\Lambda_{B_j}(s)| \,\,\,\,\,\,\forall
\,\,s \geq 0$
\end{lemma}
\begin{proof}
Let $c_{ij}(s)$ be the number of active tokens starting in $A_i$ and coalescing with active tokens starting in  $A_j$. Then,
\begin{equation}
|\Lambda_B(s)| = \sum_{j=1}^{m(t)} |B_j| - \sum_{i=1}^{m(t)} \sum_{j=1}^{m(t)} c_{ij}(s) \leq \sum_{j=1}^{m(t)}(|B_j| - c_{jj}(s)) =\sum_{j=1}^{m(t)} |\Lambda_{B_j}(s)|
\end{equation}
\end{proof}

Now using the Markov property we can upperbound the number of active tokens at any time as follows. In the beginning all the nodes of the graph $G=(V,E)$ are active. Hence we need to analyze $N(t) = E\left [ |\Lambda_V(t)| \right ]$. Suppose $O_i \subset V$ be an arbitrary subset of $V$. Since $V$ is finite the collection of all subsets, $\{O_i,\,i \in {\cal I}\}$, can be indexed by a finite index set ${\cal I}$. Denote $O_{ij} = A_i \cap O_j$. It follows that,
\begin{eqnarray} \label{ineq:xi_rs}
N(t+s) &=& E\left [|\Lambda_V(t+s)|\right ]\\ \nonumber &= &E \left [ E\left [|\Lambda_V(t+s)| \mid
\Lambda_V(t) \right ] \right ]=\sum_{j \in {\cal I}}  Prob(\Lambda_V(t)=O_j)E\left [|\Lambda_V(t+s)|
\mid \Lambda_V(t)=O_j \right ]
\\ \nonumber &\stackrel{(a)}{=}& \sum_{j \in {\cal I}}  Prob(\Lambda_V(t)=O_j)E\left [|\Lambda_{O_j}(s)| \right ]
 \\ \nonumber &\stackrel{(b)}{\leq}& \sum_{j \in {\cal I}}  Prob(\Lambda_V(t)=O_j) \left ( \sum_{i=1}^{m(t)} E\left [ |\Lambda_{O_j \cap A_i}(s)| \right ] \right )  = \sum_{j \in {\cal I}}  Prob(\Lambda_V(t)=O_j) \left ( \sum_{i=1}^{m(t)} E\left [ |\Lambda_{O_{ij}}(s)| \right ] \right )
\end{eqnarray}
where, (a) follows from Markovianity, (b) follows from Lemma~\ref{lem.icoal}.
Next, since for all $i,j$ we have $O_{ij} \subset A_i\subset V$ we can apply Lemma~\ref{lem:car_bound} and obtain:
\begin{eqnarray*}
\sum_{i=1}^{m(t)} E[|\Lambda_{O_{ij}}(s)|] &\leq& \sum_{i=1}^{m(t)} (|O_{ij}| -
(|O_{ij}|-1)\alpha_{s}(\ca_t))\\ &=& (1-\alpha_{s}(\ca_t)) \sum_{i=1}^{m(t)} |O_{ij}| +
m(t)\alpha_s(\ca_t) = (1-\alpha_s(\ca_t)) |O_{j}| +
m(t)\alpha_s(\ca_t)
\end{eqnarray*}
Substituting this result in the inequality~(\ref{ineq:xi_rs}) we
obtain
\begin{eqnarray} \nonumber
N(t+s) = E[|\Lambda_V(t+s)|] &\leq& (1-\alpha_s(\ca_t)) \sum_{j \in {\cal I}}  Prob(\Lambda_V(t)=O_j) |O_{j}| +
m(t)\alpha_s(\ca_t)\\ \label{eq.iterate} &&= (1-\alpha_s(\ca_t)) E[|\Lambda_V(t)|] + m(t)\alpha_s(\ca_t)\\ \nonumber && \stackrel{(a)}{\leq} \left (1-{\alpha_s(\ca_t)\over 2}\right ) E[|\Lambda_V(t)|]  \leq  \exp\left (-{\alpha_s(\ca_t)\over 2}\right ) N(t)
\end{eqnarray}
where $(a)$ follows from the choice of the number of partitions that it is one half of the number of active tokens at time $t$. To prove Equation~\ref{e.finalcontrac} we note from Equation~\ref{eq.iterate} that for any $r$ and $s$ such that $t \leq r \leq r+s \leq 2t$ we have,
\begin{eqnarray*}
 N(r+s) &\leq& (1-\alpha_s(\ca_t)) N(r) + m(t)\alpha_s(\ca_t) \leq (1-\alpha_s(\ca_t)) N(r) + {\alpha_s(\ca_t) \over 2} N(2t) \\ && \leq (1-\alpha_s(\ca_t)) N(r) + {\alpha_s(\ca_t) \over 2} N(r) \leq \left (1-{\alpha_s(\ca_t) \over 2} \right ) N(r) \leq  \exp\left (-{\alpha_s(\ca_t)\over 2}\right ) N(r)
\end{eqnarray*}
where the third inequality follows from the fact that since $r \leq 2t$ we have $N(2t) \leq N(r)$. Now iterating over $s$ $\left \lfloor {t \over s} \right \rfloor$ Equation~\ref{e.finalcontrac} follows.
\section{Proof of Theorem~\ref{thm:mainresult}}
We first consider the case where
\begin{equation} \label{e.slowcase}
2 \leq {N(t) \over 4} \leq {N(2t) \over 2}
\end{equation}
This implies that $N(t) \geq 8$ and $2 N(2t) \geq N(t)$.  If these assumptions are violated then we are in the case where either $N(t) \leq 8$ or
$$
N(2t) \leq {1 \over 2} N(t)
$$
First, consider the situation when Equation~\ref{e.slowcase} is satisfied. We will choose the collection $\ca_t$ and the time step $s$ so that assumptions underlying Equation~\ref{e.finalcontrac} are satisfied. Specifically, we let $\ca_t$ be the collection of balls $B(u,R_t)$ of radius $R_t$ for suitable vertices $u \in V$ to cover the graph $G$. We select radius, $R_t$ as follows:
$$R_t = \sqrt{8n\over C_0N(t)},\,\,s=s_t = R_t^2$$
where $C_0$ is the constant satisfying Equation~\ref{e.quadgrowth}. We can assume that, $s \leq t/2$. This is because if this condition is violated then, we have
\begin{equation} \label{ssmt}
N(t) \leq {128 n \over C_0 t}
\end{equation}
which satisfies the condition of the Theorem and there is nothing to prove.

So we suppose $s \leq t/2$. The number of partitions, $$m(t) \leq  \left \lfloor {n \over C_0 R_t^2} \right \rfloor + 1 \leq \left \lfloor {N(t) \over 8} \right \rfloor+ 1 \leq \left \lfloor {N(t) \over 4} \right \rfloor.$$
This ensures that assumptions underlying Equation~\ref{e.finalcontrac} are satisfied. Consequently, we get
$$
N(2t) \leq N(t) \exp\left ( - \left ( {t \over 2 R_t^2} \right ) {C_2 \over \log(R_t)} \right )
$$
Denoting $$f_t = {N(t) \over n} {t \over \log(t)},\,\,t \geq 2$$ and substituting for $$R_t = \sqrt{8n\over C_0N(t)} = \sqrt{8 t \over C_0 f_t \log(t)}$$ we get,
\begin{eqnarray*}
f_{2t} &\leq &f_t { 2 \log(t) \over \log(2t)} \exp\left ( - \left ( {t \over 2 R_t^2} \right ) {C_2 \over \log(R_t)} \right ) \leq f_t \exp\left ( \log(2) - \left ( {t \over 2 R_t^2} \right ) {C_2 \over \log(R_t)} \right ) \\
&&\leq  f_t \exp\left (\log(2) - \left ( {C_0 f_t \log(t) \over 16} \right ) {C_2 \over 0.5 \log({8t \over C_0 f_t \log(t)})} \right ) \\ &&\leq f_t \exp\left (\log(2) - {C_0C_2 \over 8}  f_t { \log(t) \over \log({8\over C_0})- \log (f_t) +  \log({t \over \log(t)})} \right )
\end{eqnarray*}
where the second inequality follows from the fact that $\log(t)/\log(2t) \leq 1$. Now we note that $f_t \leq t/\log(t)$. Consequently, if ${8\over C_0} < f_t$ then $${\log(t) \over \log({8\over C_0})- \log (f_t) +  \log({t \over \log(t)})} \geq 1.$$
Also simultaneously if $f_t \geq {8\log(2) \over C_0C_2}$ we get $f_{2t} \leq f_t$. On the other hand if any of these conditions are violated we get $$f_{2t} \leq 2 \max( { 8 \over C_0}, {8\log(2) \over C_0C_2}) \stackrel{\Delta}{=} \tilde C.$$
This implies that,
$$
f_{2t} \leq \max \left (\tilde C , f_t \right )  \implies N(2t) \leq \max \left ( \tilde C {n \log(2t) \over 2t}, {N(t) \over 2}(1 + 1/\log(t)) \right )
$$
Finally, we have two cases to consider: (1) if Equation~\ref{e.slowcase} itself is violated we have, $N(2t) \leq {1 \over 2} N(t)$; (2) Equation~\ref{ssmt} holds. In all of these cases Eq.~\ref{eq:strngbnd} is satisfied and the result follows.
\subsection{Proof of Lemma~\ref{lem:tsteptrans}}
\begin{proof}
The proof follows directly along the lines of Pettarin et al \cite{pettarin} (Lemma 9). We provide a brief sketch of their proof here for the sake of completion. First, let $P_t(x,y)$ denote the probability that a walk starting at node $x$ at time zero is at node $y$ at time $t$. Now consider two walks, one starting at node $u$ and another starting at node $w$ at time zero. Note that the two walks are independent and they have their own corresponding transition probabilities. Let $N(u,w,T_0)$ be the mean number of times the two walks starting at $u$ and $w$ meet in the time interval $[0,T_0]$. Then noting that the two walks could meet at any node $v$ at any time $t \in [0,\,T_0]$ we obtain,
$$
N(u,w,T_0) = \sum_{t=0}^{T_0} \sum_v P_t(u,v)P_t(w,v)
$$
This is because $P_t(u,v)P_t(w,v)$ is the probability that both walks starting at $u$ and $w$ respectively are at the same node $v$ at time $t$. Summing over the different possibilities leads to the above result. Using this fact Pettarin et al establish that,
$$
\alpha_{R^2}(B(u,R)) \geq {N(u,w,R^2)\over \max_{v \in B(u,R)} N(v,v,R^2)},\,\,w\in B(u,R)
$$
Here, $N(v,v,R^2)$ is the number of times two walks starting at the same node $v$ meet again in the time interval $[0,R^2]$. The problem now boils down to lower bounding $N(u,w,R^2)$ and upper bounding $N(v,v,R^2)$. We are now ready to substitute the Gaussian t-step bounds to establish the result. Specifically, let
$$
D = \{v \in V \mid d(v,u) \leq 2R,\,d(v,w) \leq 2R\}
$$
We also note that since $B(u,R) \subset D$ and the graph satisfies the geometric neighborhood property we have $|D| \geq C_0 R^2$. So
\begin{eqnarray*}
N(u,w,R^2) &=& \sum_{t=0}^{R^2} \sum_v P_t(u,v)P_t(w,v) \geq \sum_{t=R^2/2+1}^{R^2} \sum_{v \in D} P_t(u,v)P_t(w,v) \\ &&\geq \sum_{t=R^2/2+1}^{R^2}  \sum_{v \in D} \left ( {C_3 \over t}\right )^2 \exp( -{d^2(v,u) + d^2(v,w) \over C_4t} )
\end{eqnarray*}
By bounding $d^2(v,u)$ and $d^2(v,w)$ with $4R^2$ we obtain $N(u,w,R^2) = \Omega(1)$. Next we use the fact that there are no more than $C_1 k \Delta$ nodes in any annulus of size $\Delta$ at distance $k$ to obtain an upper bound for $N(v,v,R^2)$. Specifically, by taking an annulus of size one, our geometric condition implies that there are no more than $C_1R$ nodes at distance $R$. So,
\begin{eqnarray*}
N(v,v, T) &= &\sum_{t=0}^T \sum_x P_t(v,x)P_t(v,x) \leq 1 + \sum_{t=1}^T \sum_{k=1}^t \sum_{d(v,x)=k} P_t(v,x)P_t(v,x) \\ &&\leq 1 + \sum_{t=1}^T \sum_{k=1}^t C_1 k \left ( {C_3 \over t} \right )^2 \exp \left ({-2k^2 \over t}\right )
\end{eqnarray*}
The computation of the above sum follows along the same lines as in Pettarin et al (Lemma 9). It follows that $N(v,v,T) = O(\log(T))$ which is $O(\log(R))$ for $T = R$. Consequently, there is a constant $C_2$ such that,
$$
\alpha_{R^2}(B(u,R)) \geq {N(u,w,R^2)\over \max_{v \in B(u,R)} N(v,v,R^2)} = {C_2 \over \log(R)},\,\,w\in B(u,R)
$$
\end{proof}


\end{document}